\documentclass[prx,longbibliography,twocolumn]{revtex4-1}
\usepackage{graphicx}
\usepackage{comment}
\usepackage{amsmath,amssymb,amsthm} 
\usepackage{verbatim}
\usepackage{float}
\newcommand{\br}{{\bf r}}

\begin{document}

\title{What constitutes a simple liquid?}
\author{Trond S. Ingebrigtsen, Thomas B. Schr{\o}der, and Jeppe C. Dyre}
\affiliation{DNRF centre  ``Glass and Time,'' IMFUFA, Department of Sciences, Roskilde University, Postbox 260, DK-4000 Roskilde, Denmark}
\date{\today}

\begin{abstract}
Simple liquids are traditionally defined as many-body systems of classical particles interacting via radially symmetric pair potentials. We suggest that a simple liquid should be defined instead by the property of having strong correlation between virial and potential energy equilibrium fluctuations in the $NVT$ ensemble. There is considerable overlap between the two definitions, but also some notable differences. For instance, in the new definition simplicity is not a property of the intermolecular potential because a liquid is usually only strongly correlating in part of its phase diagram. Moreover, according to the new definition not all simple liquids are atomic (i.e., with radially symmetric pair potentials) and not all atomic liquids are simple. The main part of the paper motivates the new definition of liquid simplicity by presenting evidence that a liquid is strongly correlating if and only if its intermolecular interactions may be ignored beyond the first coordination shell (FCS). This is demonstrated by $NVT$ simulations of structure and dynamics of atomic and molecular model liquids with a shifted-forces cutoff placed at the first minimum of the radial distribution function. The liquids studied are: inverse power-law systems ($r^{-n}$ pair potentials with $n=18, 6, 4$), Lennard-Jones (LJ) models (the standard LJ model, two generalized Kob-Andersen binary LJ mixtures, the Wahnstr{\"o}m binary LJ mixture), the Buckingham model, the Dzugutov model, the LJ Gaussian model, the Gaussian core model, the Hansen-McDonald molten salt model, the Lewis-Wahnstr{\"o}m OTP model, the asymmetric dumbbell model, and the rigid SPC/E water model. The final part of the paper summarizes most known properties of strongly correlating liquids, showing that these are simpler than liquids in general. Simple liquids as defined here may be characterized (1)  {\it chemically} by the fact that the liquid's properties are fully determined by interactions from the molecules within the FCS, (2) {\it physically} by the fact that there are isomorphs in the phase diagram, i.e., curves along which several properties like excess entropy, structure, and dynamics, are invariant in reduced units, and (3) {\it mathematically} by the fact that the reduced-coordinate constant-potential energy hypersurfaces define a one-parameter family of compact Riemannian manifolds. No proof is given that the chemical characterization follows from the strong correlation property, but it is shown to be consistent with the existence of isomorphs in strongly correlating liquids' phase diagram. Finally, we note that the FCS characterization of simple liquids calls into question the basis for standard perturbation theory, according to which the repulsive and attractive forces play fundamentally different roles for the physics of liquids.

\end{abstract}

\maketitle

\section{Introduction}

Going back to Plato, classification or categorization is the epistemological process that groups objects based on similar properties \cite{plato}. Having primarily biological examples in mind, Aristotle defined categories as discrete entities characterized by properties shared by their members \cite{arist}. Locke in 1690 distinguished between the {\it nominal} and the {\it real essence} of an object \cite{locke}. The nominal essence comes from experience and represents the object's appearance, the real essence represents the object's deeper, constituting features. For instance, the real essence of a material thing is its atomic constitution, because this is the causal basis of all the thing's observable properties \cite{wiki}. A scientific classification is particularly useful if it reflects the {\it real} essence of the objects in question by identifying their underlying common features, from which the more obvious and easily observable {\it nominal} properties follow. Having in mind Locke's concept of real essence, we argue below for a new definition of the class of simple liquids.

Physicists love simple systems. This reflects the fundamental paradigm that in order to capture a given phenomenon, simpler is better. Most classifications in physics are clear-cut, for example the classification of elementary particles into baryons and leptons, whereas classifications in other sciences usually have a wealth of borderline cases. Due to the diversity of molecules it is reasonable to expect a definition of ``simple liquids'' to be of the latter type.

The concept of a simple liquid is old, but it remains central as evidenced by the 2003 book title ``Basic concepts for simple and complex liquids'' \cite{barrat} or the review of non-simple liquids entitled ``Theory of complicated liquids'' from 2007 \cite{kir07}. Generations of liquid-state theorists were introduced to the topic by studying Hansen and McDonald's textbook ``Theory of simple liquids'' \cite{han05}. This book first appeared in 1976 following a period  of spectacular progress in the theory of liquids, catalyzed by some of the very first scientific computer simulations.

In Ref. \onlinecite{han05} a simple liquid is defined as a classical system of approximately spherical, nonpolar molecules interacting via pair potentials. This and closely related definitions of liquid simplicity have been standard for many years \cite{fis64,ric65,tem68,ail80,gubbins}. In this definition simple liquids have much in common with the chemists' ``nonassociated liquids'' \cite{chandler}, but there are some differences. Chemists generally regard a liquid as simple even if it consists of elongated molecules, as long as these are without internal degrees of freedom and interact primarily via van der Waals forces. Many physicists would probably disagree. Thus it is far from trivial to ask: What characterizes a ``simple liquid''? More accurately: Given a classical system of rigid bodies with potential energy as a function of the bodies' centers-of-masses and their spatial orientations, is it possible to give a quantitative criterion for how simple the system is? If yes, is simplicity encoded uniquely in the potential energy function or may the degree of simplicity vary throughout the phase diagram?

Recent works identified and described the properties of ``strongly correlating liquids'' \cite{ped08,ped10,effectivetemperature,baileyOrderParameters,paperI,paperII,paperIII,paperIV,paperV,paper00,sch09,ped11}. In these liquids, by definition, the virial $W$ and the potential energy $U$ correlate strongly in their constant-volume thermal-equilibrium fluctuations. Recall that the average virial $\langle W\rangle$ gives the contribution to pressure from intermolecular interactions, added to the ideal-gas term $Nk_BT$ deriving from momentum transport via particle motion (below $p$ is the pressure, $V$ the volume, $N$ the number of particles, $k_B$ Boltzmann's constant, and $T$ the temperature):

\begin{equation}
pV
\,=\,Nk_BT+\langle W\rangle\,.
\end{equation}
The term ``strongly correlating liquid'' refers to the case when the $WU$ correlation coefficient in the $NVT$ ensemble is larger than $0.9$ \cite{paperI}. If angular brackets denote an $NVT$ ensemble average, the correlation coefficient $R$ is defined by

\begin{equation}\label{Rdef}
R
\,=\,\frac{\langle \Delta W \Delta U \rangle}{\sqrt{\langle (\Delta W)^{2} \rangle \langle (\Delta U)^{2} \rangle}}\,.
\end{equation}
An example of a strongly correlating liquid is the standard Lennard-Jones liquid at typical condensed-phase state points, i.e., not far from the solid-liquid coexistence line. Many other systems, including some molecular models, have been shown to be strongly correlating; we refer the reader to papers that derive and document the several simple properties of strongly correlating liquids 
\cite{ped08,ped10,effectivetemperature,baileyOrderParameters,paperI,paperII,paperIII,paperIV,paperV,paper00,sch09,ped11}, reviewed briefly in Ref. \onlinecite{ped11}. These properties are summarized in Sec. \ref{prop} after the presentation of the simulation results.

The present work is motivated by developments initiated by recent findings by Berthier and Tarjus \cite{wcabertier1,wcabertier2}. These authors showed that for the viscous Kob-Andersen binary Lennard-Jones mixture \cite{ka1,ka2} the dynamics is not  reproduced properly by cutting the potentials at their minima according to the well-known Weeks-Chandler-Andersen (WCA) recipe \cite{wca}. The role of the cutoff was subsequently studied in two papers, showing that placing a shifted-forces cutoff at the first minimum of the pair correlation function gives good results for Lennard-Jones type systems \cite{FCS1,FCS2}. This applies not only at moderate densities, but also at very high densities. Applying the same cutoff to water does not work very well \cite{pal06}. Water is an example of a non-strongly correlating liquid with $R\approx 0$ at ambient conditions, a consequence of water's density maximum \cite{paperI}. These findings led us to speculate whether it is a general property of strongly correlating liquids that the intermolecular interactions may be ignored beyond the FCS without compromising accuracy to any significant extent. The main part of the present paper shows that, indeed, using such an ``FCS cutoff'' gives accurate simulation results if and only if the liquid is strongly correlating.

The paper presents results obtained from computer simulations of 15 different systems, some of which are strongly correlating. We investigate the role of the FCS in determining liquid structure and dynamics. Structure is probed by the radial distribution function (RDF), dynamics by the incoherent or coherent intermediate scattering function (ISF) at the wavevector of the static structure factor maximum. The numerical evidence is clear. By varying the cutoff of the intermolecular forces, we find that in order to get accurate simulation results it is enough to take into account the interactions within the FCS {\it if and only if} the liquid is strongly correlating. In other words, for strongly correlating liquids interactions beyond the FCS are unimportant, and this applies {\it only} for these liquids. At present there is no rigorous argument for this empirical ``FCS property'', but we show that it is consistent with known properties of strongly correlating liquids.

The FCS property of strongly correlating liquids documented below emphasizes further that these are simpler than liquids in general. A number of other simple properties of strongly correlating liquids were reported previously \cite{ped08,ped10,effectivetemperature,baileyOrderParameters,paperI,paperII,paperIII,paperIV,paperV,sch09,ped11}. Altogether, these facts motivate our new definition of liquid simplicity.

Section \ref{sim_sec} presents the results from molecular dynamics simulations and Sec. \ref{sum} summarizes the results. Section \ref{SCL_sec} gives an overview of the many simple properties of strongly correlating liquids, motivating our suggestion that a liquid is to be defined as simple whenever it is strongly correlating at the state point in question. Section \ref{disc} gives a few concluding remarks.

\section{Molecular dynamics simulations of atomic and molecular liquids}\label{sim_sec}

In a computer simulation the intermolecular interactions, which usually extend in principle to infinity, are truncated at some cutoff distance $r_{c}$ beyond which they are ignored \cite{FCS1}. To avoid a discontinuity in the force, which can severely affect the simulation results \cite{FCS1,tildesley}, the simulations reported below use potentials truncated such that the force goes continuously to zero at $r_{c}$. This is done by applying a so-called shifted forces (SF) cutoff \cite{tildesley,kan85,hal08} where, if the pair potential is $v(r)$ and the pair force is $f(r)=-v'(r)$, the shifted force is given by 

\begin{equation}
f_{\rm SF}(r)\,=\,
\begin{cases}  f(r)-f(r_c) & \text{if}\,\, r<r_c\,, \\ 
0 &\text{if}\,\, r>r_c\,.
\end{cases} 
\end{equation}
This corresponds to using the following pair potential below $r_c$: $v_{\rm SF}(r) = v(r) - v'(r_c) ( r-r_c) - v(r_c)$. Using an SF cutoff gives more accurate results and better numerical stability than using the standard shifted-potential (SP) cutoff \cite{FCS1}. This is so despite the fact that an SF cutoff does not have the correct pair force for any $r$, whereas the pair force is correct below $r_c$ for an SP cutoff. Apparently, avoiding discontinuity of the force at $r_c$ is more important than maintaining the correct force. It was recently discussed why adding a linear term to the pair potential affects neither structure nor dynamics to any significant extent \cite{paperII}. The reason is that, when one nearest-neighbor distance decreases, others increase in such a way that their sum is virtually constant. This argument is exact in one dimension and holds to a good approximation in 3D constant-volume ensemble simulations \cite{paperII} (in constant-pressure ensemble simulations the volume fluctuates and the argument no longer applies). Coulomb interactions have also been treated by the SF cutoff procedure. Although the Coulomb interaction is long-ranged and conditionally convergent, when $r_c$ is sufficiently large an SF cutoff gives results close to those of the standard, much more involved Ewald summation method \cite{fennell,jesper}.

All simulations were performed in the $NVT$ ensemble with periodic boundary conditions and the Nose-Hoover algorithm \cite{nose,hoover,frenkel}. We used the RUMD molecular dynamics package developed in-house, which is optimized for simulations of small systems on state-of-the-art GPU hardware (a few thousand entities) \cite{rumd}. For the molecular models bond lengths were held fixed using the time-symmetrical central difference algorithm \cite{nvttoxvaerd,toxconstraintnph,toxconstraintnve}. 

In the following we investigate for several systems whether it is possible to choose an FCS cutoff, i.e., a cutoff at the first minimum of the pair correlation function, and still get the correct physics. We start by studying strongly correlating atomic liquids. Then data are presented for atomic liquids that are not strongly correlating. Finally data are given for two strongly correlating molecular liquids and a water model. Details of the models studied, number of particles, etc, are given in Appendix \ref{pairP}.

\subsection{Three inverse-power-law (IPL) fluids}

We consider first systems with 100\% correlation between virial and potential energy equilibrium fluctuations in the $NVT$ ensemble. It follows from the definition of the virial $W=-1/3\sum{\bf r}_i\cdot\nabla_i U$ \cite{tildesley} that a necessary and sufficient condition for $W$ to correlate perfectly with $U$ is that the potential energy is a so-called Euler homogeneous function of the particle coordinates ${\bf r}_i$. This is the case for systems with inverse power-law (IPL) pair potentials ($v(r)\propto r^{-n}$), but note that potentials with non-trivial angular dependence may also be Euler homogeneous. 

We simulated single-component IPL pair-potential systems with exponents $n = 18, 6, 4$ at density $\rho = 0.85$. Each system was simulated at two temperatures. The simulated systems range from very harsh repulsive ($n=18$) to quite soft and long ranged ($n=4$). Each system was simulated with four SF cutoffs. The role of the cutoff is investigated by choosing three different cutoffs: one placed at the first minimum of the RDF, one corresponding to the half height of the RDF from its first maximum to the first minimum, and one placed to the right of the RDF first minimum displaced the same amount as the difference between the first and the second cutoff. 

\begin{figure}[H]
  \centering
  \includegraphics[width=70mm]{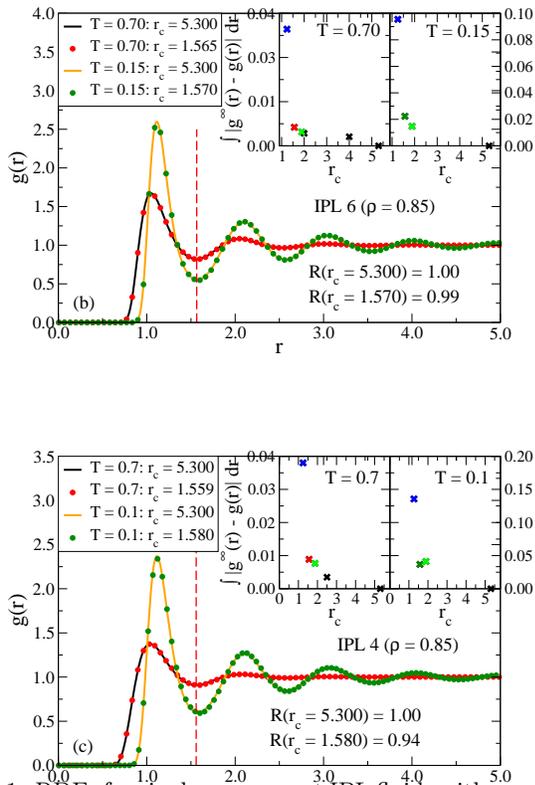}
  \vspace{-20pt}
  \caption{RDFs for single-component {IPL} fluids with exponents $n=18,6,4$, each simulated at two temperatures at density $\rho = 0.85$. The black and orange curves show reference simulation results with large cutoffs representing the ``true'' IPL behavior, the red and green dots give results from simulations with an FCS cutoff (marked by the vertical red dashed lines). The insets quantify the deviations in the RDF from the reference RDF as functions of the cutoff; deviations increase dramatically when the cutoff enters the FCS (blue crosses). In each subfigure the virial/potential-energy correlation coefficient $R$ is given for the simulated cutoff (this quantity is exactly unity for IPL systems with infinite cutoff).}
  \label{IPLgrs}
\end{figure}

The RDFs $g(r)$ are shown for $n = 18, 6, 4$ in Fig. \ref{IPLgrs}; $n=12$ gives similar results (not shown). The simulations with an SF cutoff at the first minimum of the RDF -- referred to as FCS-cutoff simulations -- give a faithful representation of the structure. The insets show the deviations in RDF as functions of the cutoff between results for an SF cutoff and ``true'' results, quantified by integrating the numerical difference in the pair correlation function. Clearly, deviations increase sharply when the cutoff enters the FCS (blue crosses). 

We simulated also the $n=3$ and $n=1$ IPL fluids (results not shown), the latter is usually termed the one-component plasma. For both systems an FCS cutoff does not lead to the correct physics. This is most likely because the potentials are too long ranged for such a short cutoff to make good sense. In fact, both are so long ranged that they do not have a proper thermodynamic limit for which the exponent must be larger than the dimension \cite{fis66}. An indication that an FCS cutoff works poorly when the IPL exponent approaches the dimension is seen for the $n=4$ simulation, for which the $WU$ correlation coefficient for the FCS system starts to deviate significantly from unity. Moreover, almost invisible in the figure is the fact that the $n=4$ pair correlation function's first maximum deviates slightly between FCS and true simulations.

\begin{figure}[H]
  \centering
  \includegraphics[width=70mm]{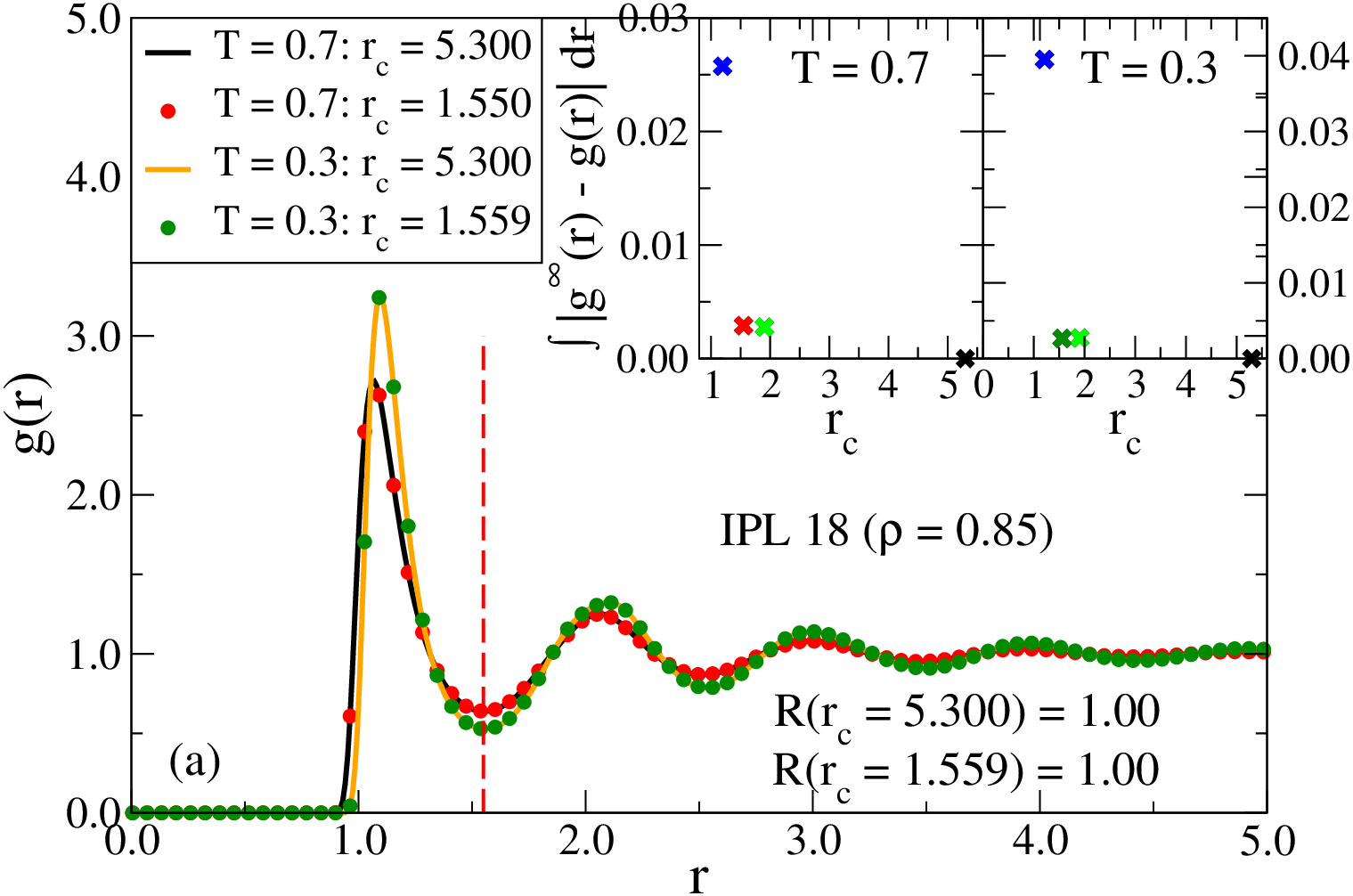}
  \vspace{-20pt}
\end{figure}  
\begin{figure}[H]
  \centering
  \includegraphics[width=70mm]{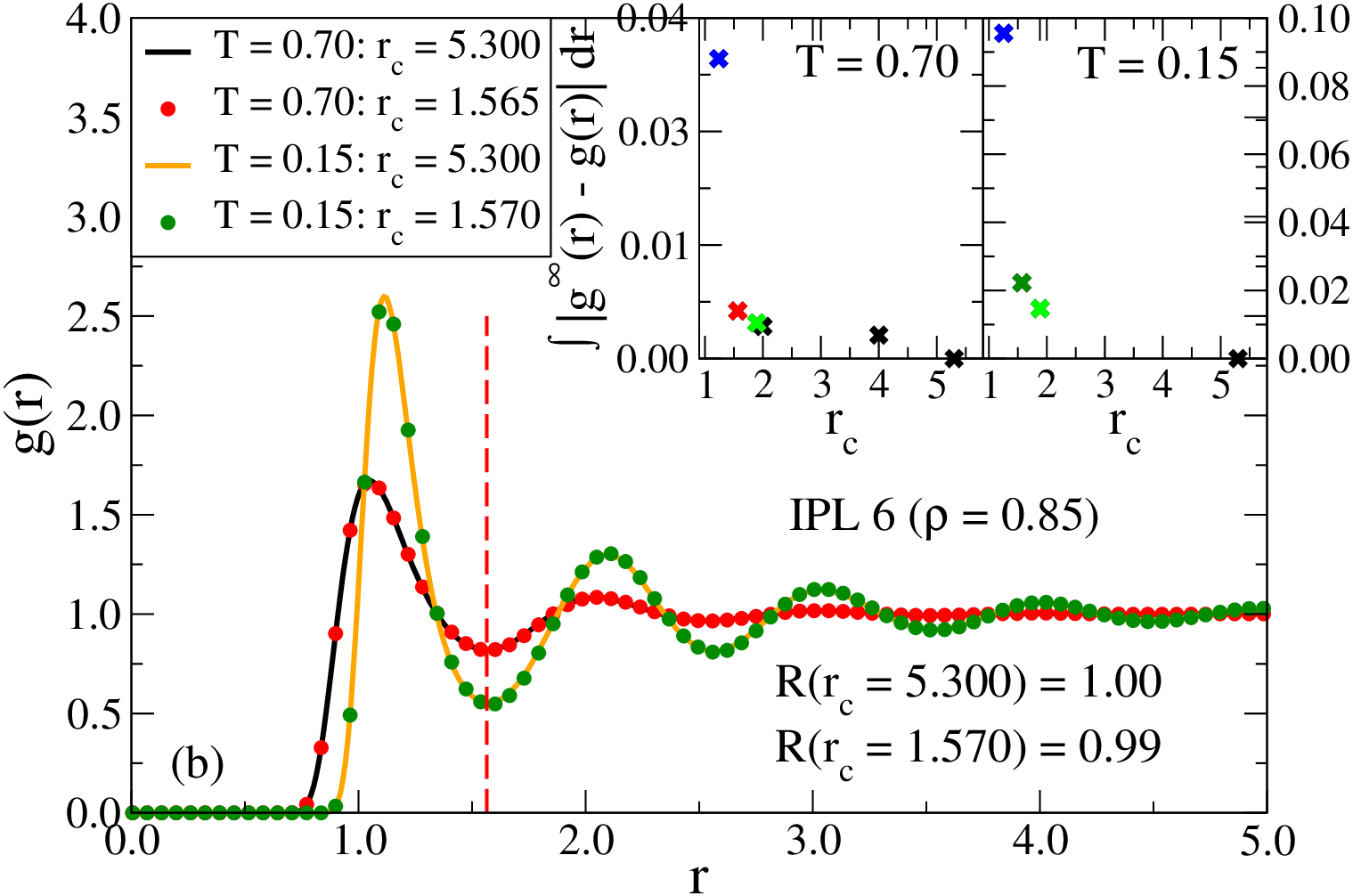}
  \vspace{-20pt}
\end{figure}    

\begin{figure}[H]
  \centering
  \includegraphics[width=70mm]{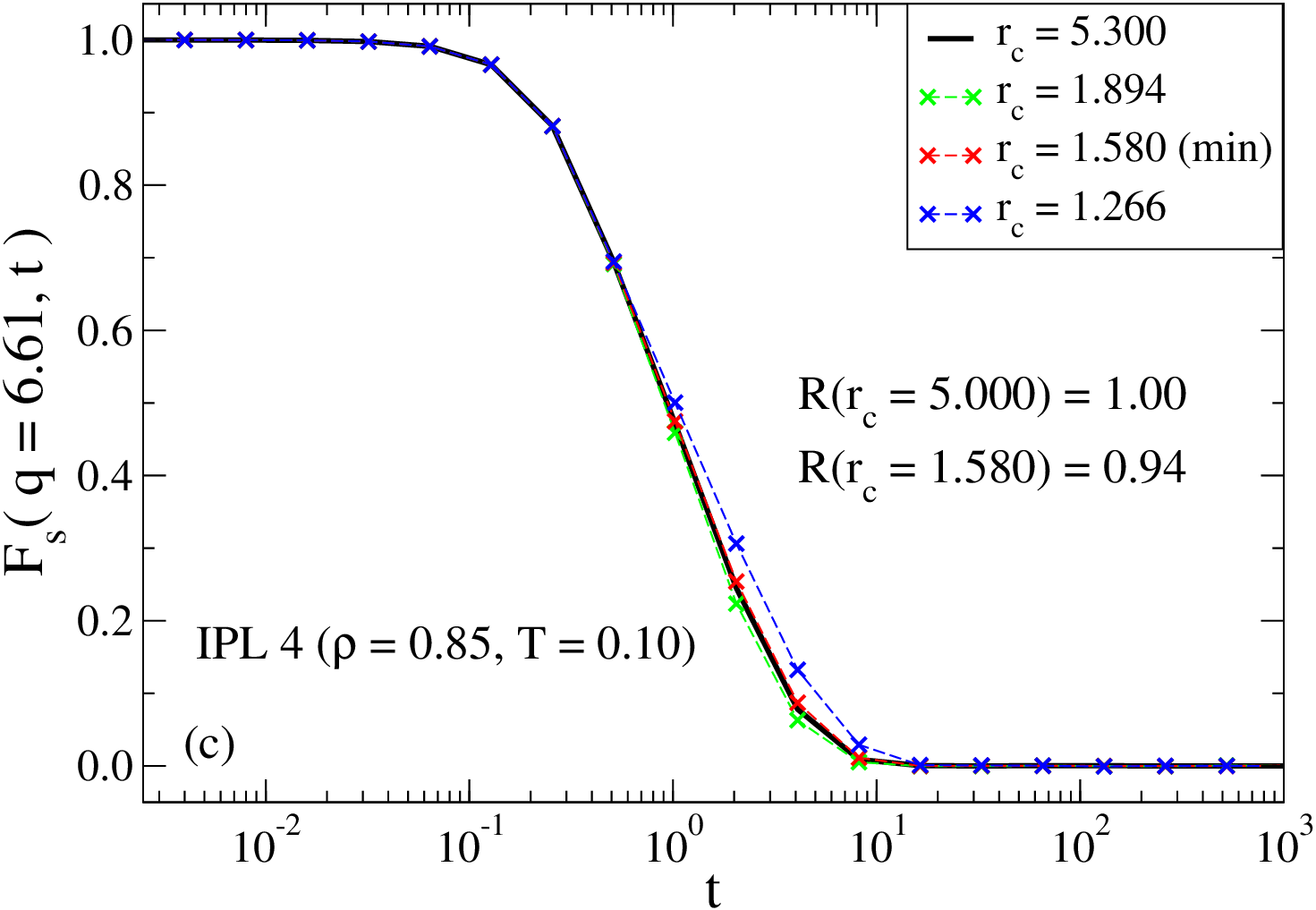}
  \vspace{-20pt}
  \caption{Incoherent intermediate scattering functions (ISFs) for the {IPL} fluids at the lowest-temperature state points of Fig. \ref{IPLgrs}.
    The black curves give results for a large cutoff, the red crosses for an FCS cutoff (marked by the vertical red dashed lines in Fig. \ref{IPLgrs}).
    (a) $n=18$,  $T = 0.30$; 
    (b) $n=6$,   $T = 0.15$;
    (c) $n=4$,   $T = 0.10$.
  }
  \label{IPLFsAs}
\end{figure}

Figure \ref{IPLFsAs} shows the incoherent intermediate scattering functions (ISFs) for the low-temperature state points of each of the three IPL systems evaluated at the wavevector corresponding to the first maximum of the static structure factor. A good representation of the dynamics is obtained for all systems when the FCS cutoff is used.

\begin{figure}[H]
  \centering
  \includegraphics[width=70mm]{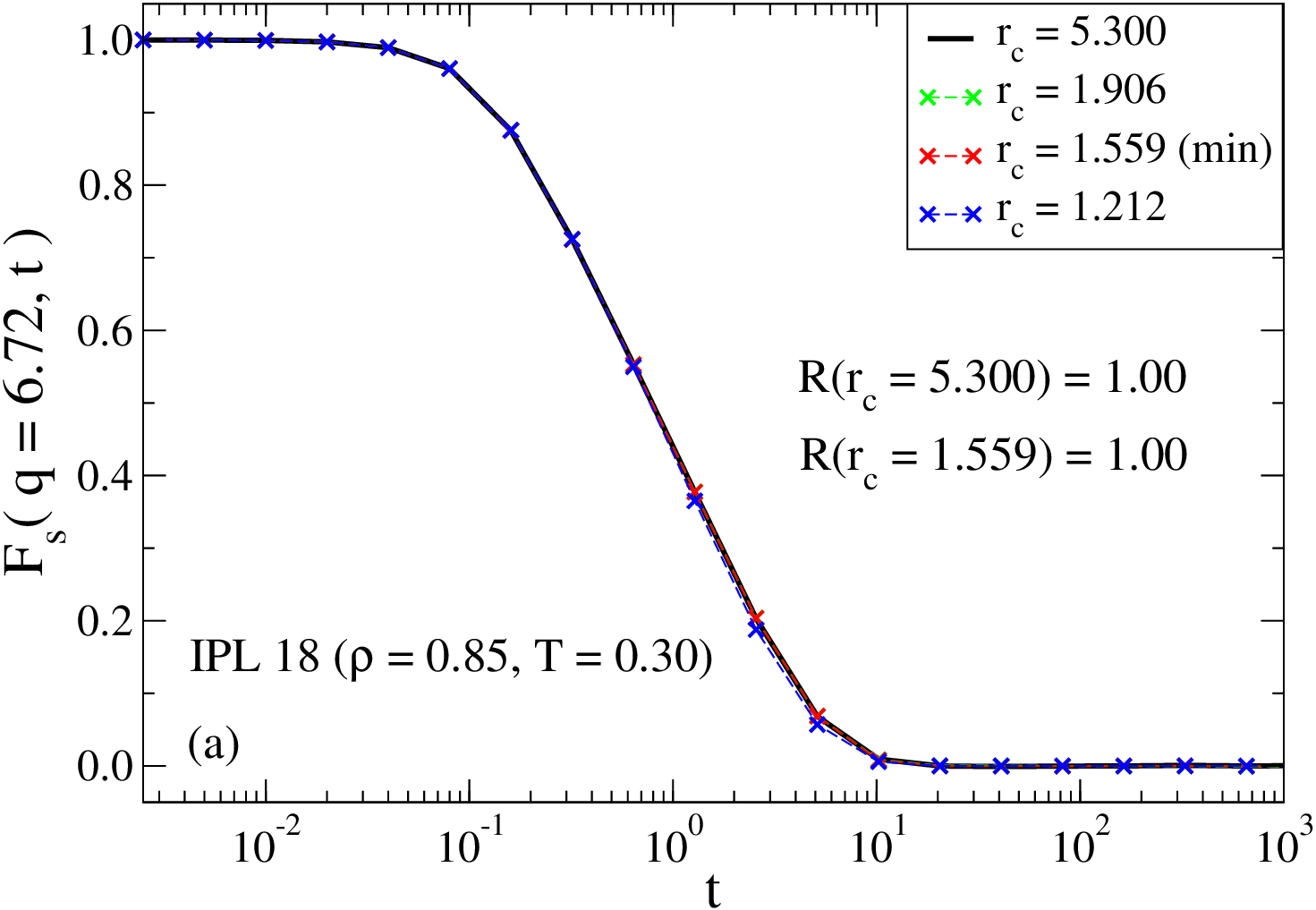}
  \vspace{-20pt}
\end{figure}  
\begin{figure}[H]
  \centering
  \includegraphics[width=70mm]{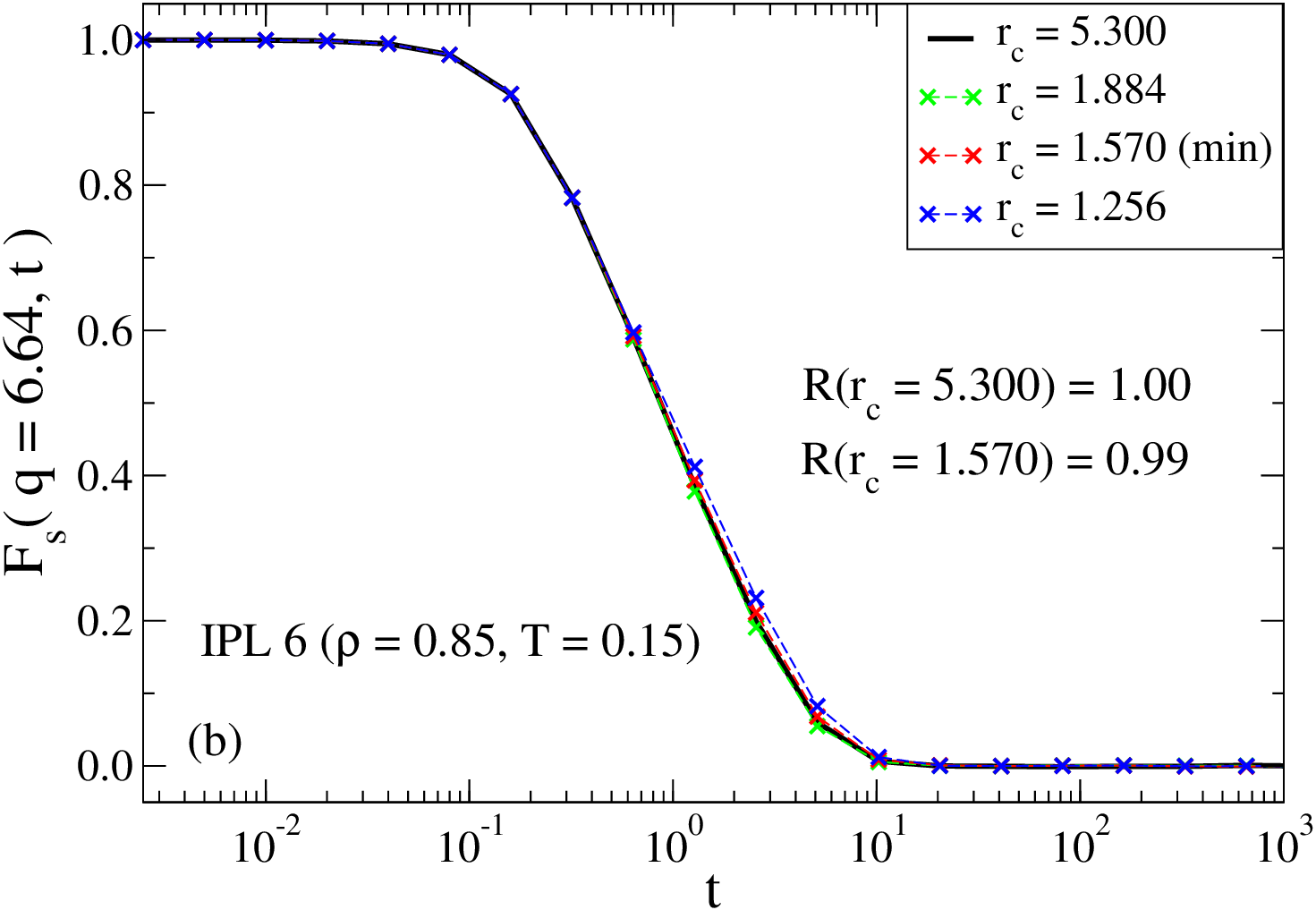}
  \vspace{-20pt}
\end{figure}    

\subsection{Lennard-Jones type liquids}

\begin{figure}[H]
  \centering
  \includegraphics[width=70mm]{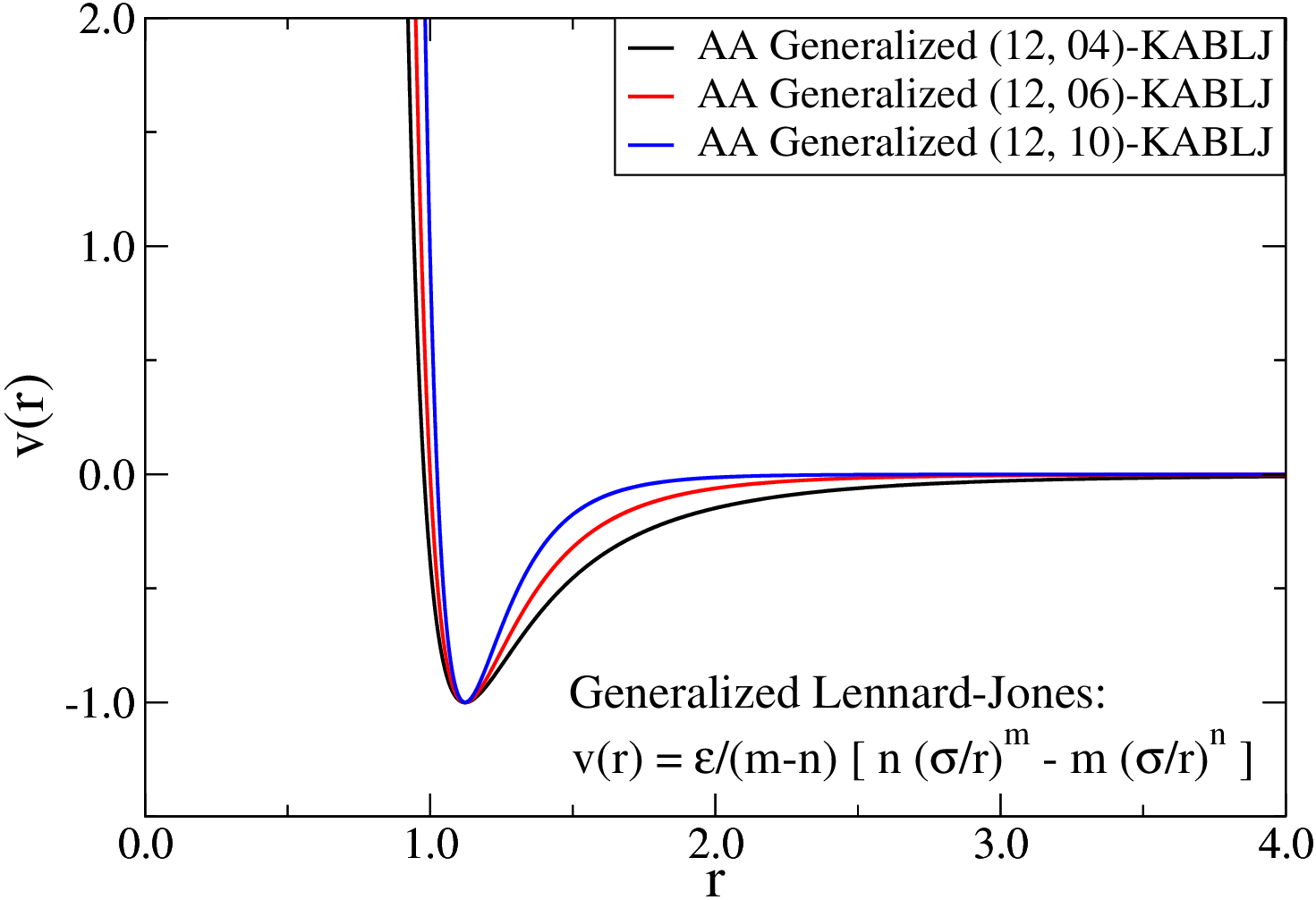}
  \caption{The AA-particle generalized Kob-Andersen (KABLJ) pair potentials with fixed repulsive exponent $12$ and three different attractive exponents $n=4,6,10$. The model parameters are: $\epsilon_{AA}=1.00$, $\epsilon_{AB}=1.50$, and $\epsilon_{BB}=0.50$, $\sigma_{AA}=1.12$, $\sigma_{AB}=0.90$, $\sigma_{BB}=0.99$, $m_A=m_B=1$.}
  \label{pairKABLJ}
\end{figure}

Next, we consider the most studied potential in the history of computer simulations, the Lennard-Jones (LJ) pair potential, 

\begin{equation}\label{LJ}
  v_{\rm LJ}(r) = 4\epsilon \Big[ \Big( \frac{\sigma}{r}\Big)^{12} - \Big(\frac{\sigma}{r}\Big)^{6} \Big]\,.
\end{equation}
Here $\sigma$ and $\epsilon$, define, respectively, the length and energy scale of the interaction (dimensionless units defined by $\sigma=\epsilon=1$ are used below). This potential does not have 100\% virial/potential-energy correlation, but has still quite strong correlations with correlation coefficients $R>0.9$ in the condensed-fluid part of the phase diagram (also in the crystalline phase \cite{paperII}). We studied the single-component LJ (SCLJ) liquid, two generalized 80/20 Kob-Andersen binary LJ (KABLJ)  mixtures with repulsive exponent $12$ and attractive exponents $n=4,10$, and the Wahnstrom 50/50 binary LJ (WABLJ) mixture (Fig. \ref{pairKABLJ} and Appendix \ref{pairP} gives model details). The influence of an SF cutoff on simulation accuracy was investigated recently for the SCLJ liquid and the standard KABLJ mixture ($n=6$) \cite{FCS1,FCS2}, but for completeness we include results for the SCLJ system here as well.

\begin{figure}[H]
  \centering
  \includegraphics[width=70mm]{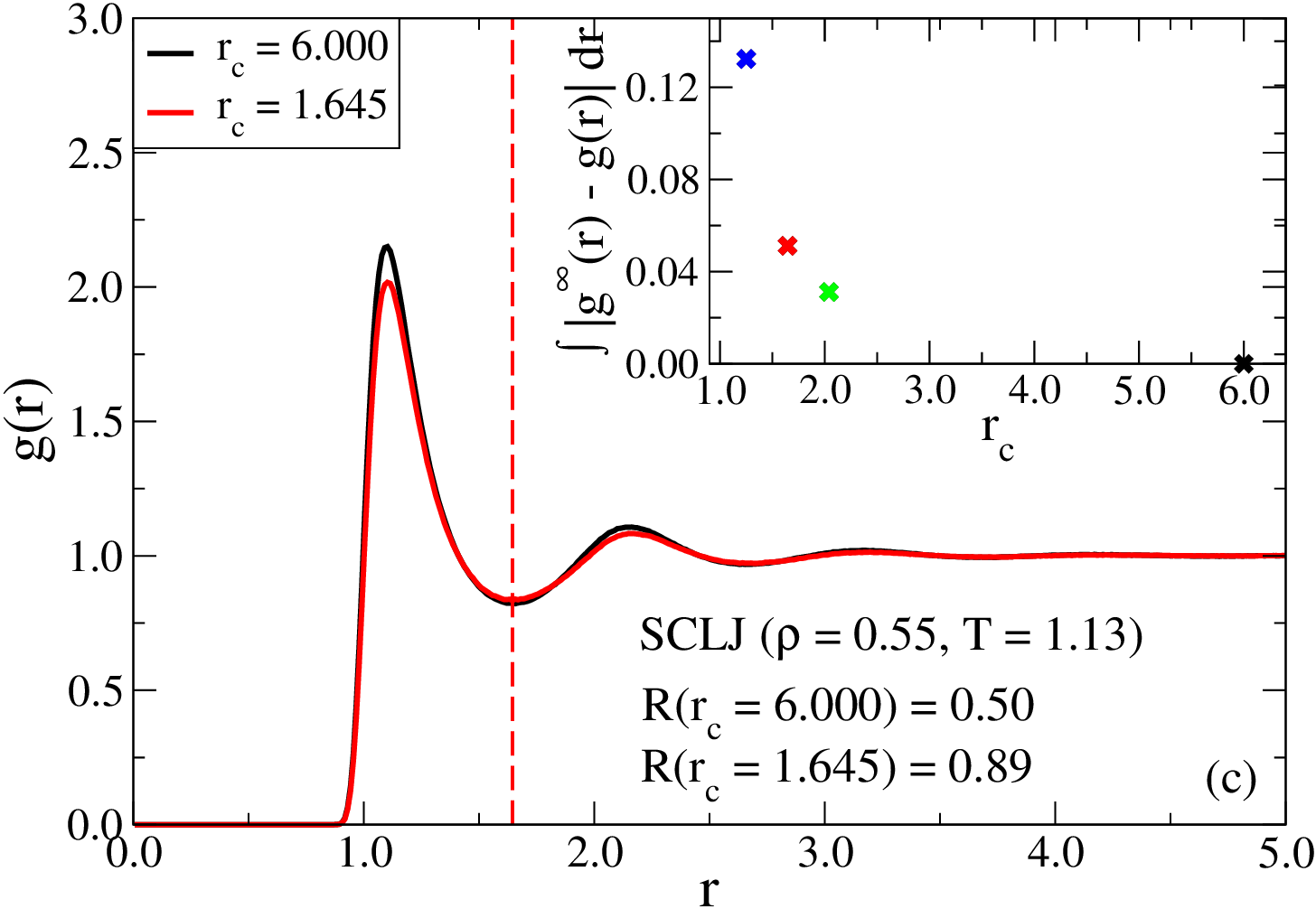}
  \caption{RDFs for the single-component Lennard-Jones (SCLJ) liquid at three different state points: 
(a) $\rho=0.85$, $T=0.70$ ($R = 0.96$); 
(b) $\rho=0.85$, $T=1.00$ ($R = 0.97$); 
(c) $\rho=0.55$, $T=1.13$ ($R = 0.50$). 
The black curves show reference simulations with large cutoffs, the red dots/curve show results from simulations with an FCS cutoff (marked by the vertical red dashed lines). The insets quantify the deviation in RDF from the reference RDF as functions of the cutoff. At all three state points deviations increase significantly when the cutoff enters the FCS (blue crosses in the insets). For state points (a) and (b), which are strongly correlating ($R>0.9$), an FCS cutoff leads to accurate results. This is not the case for state point (c) that is not strongly correlating.}
  \label{SCLJgr}
\end{figure}

\begin{figure}[H]
  \centering
  \includegraphics[width=70mm]{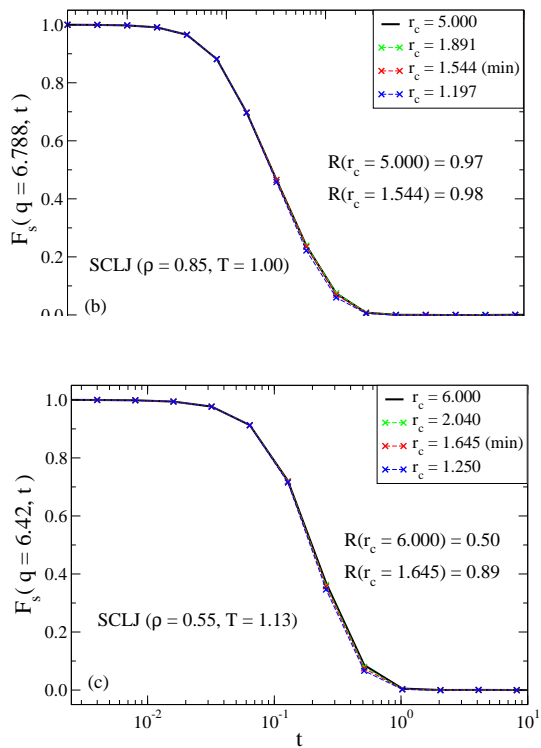}
  \caption{{ISF}s for the {SCLJ} liquid at the state points of Fig. \ref{SCLJgr}. The black curves give the reference cutoff results, the red curves the FCS cutoff results, the blue curves results for a cutoff at the half-height towards the first maximum of the RDF, the green curves results for a cutoff to the right of the minimum.  
(a) $\rho=0.85$, $T=0.70$ ($R = 0.96$); 
(b) $\rho=0.85$, $T=1.00$ ($R = 0.97$); 
(c) $\rho=0.55$, $T=1.13$ ($R = 0.50$).}
  \label{SCLJdyn}
\end{figure}

\begin{figure}[H]
  \centering
  \includegraphics[width=70mm]{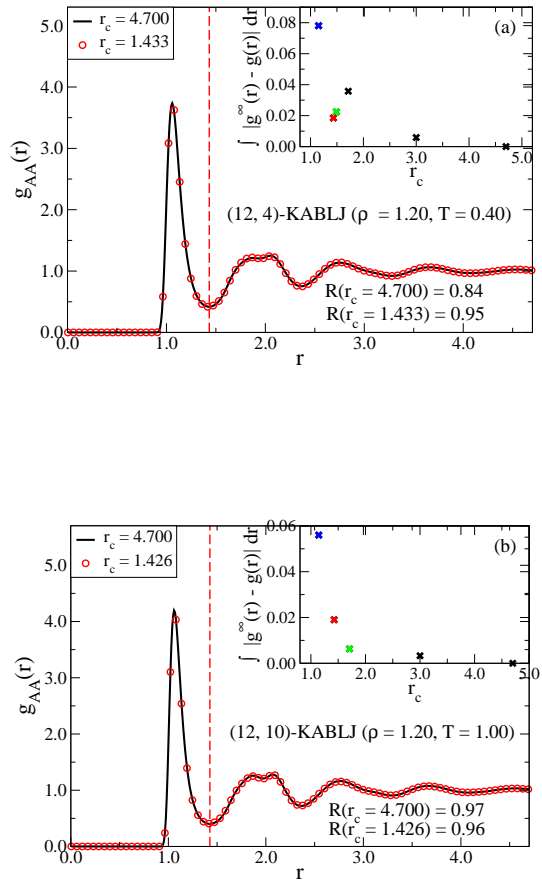}
  \caption{RDFs for generalized KABLJ mixtures with repulsive exponent $12$ and attractive exponents $n=4,10$. The black curves give the reference cutoff results, the red curves the FCS cutoff results. The insets quantify the deviation in RDF from the reference RDF as functions of the cutoff. 
(a) $n=4$, $\rho=1.20$, $T=0.40$ ($R = 0.84$); 
(b) $n=10$, $\rho=1.20$, $T=1.00$ ($R = 0.97$). }
  \label{KABLJgr}
\end{figure}

\begin{figure}[H]
  \centering
  \includegraphics[width=70mm]{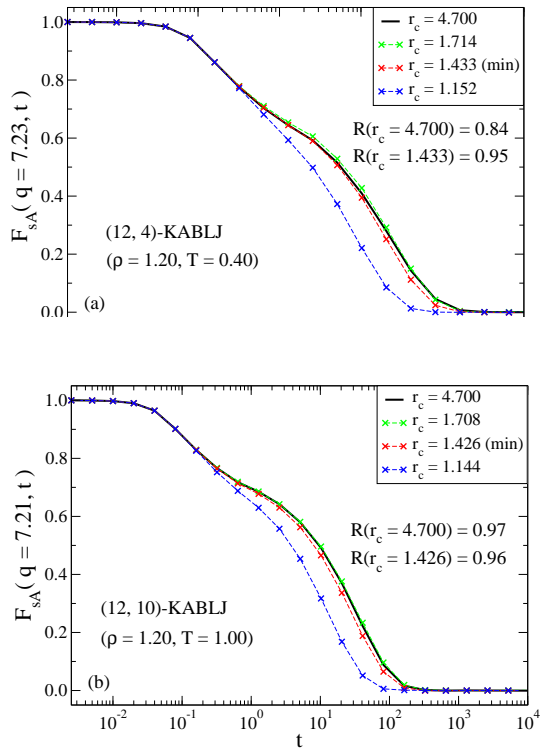}
  \caption{{ISF}s for generalized KABLJ mixtures with repulsive exponent $12$ and attractive exponents $n=4,10$.
    The red and black curves give, respectively, results for FCS cutoffs and large reference cutoffs. 
    (a) $n=4$, $\rho=1.20$, $T=0.40$ ($R = 0.84$); 
    (b) $n=10$, $\rho=1.20$, $T=1.00$ ($R = 0.97$).}
  \label{KABLJFsA}
\end{figure}

\begin{figure}[H]
  \centering
  \includegraphics[width=70mm]{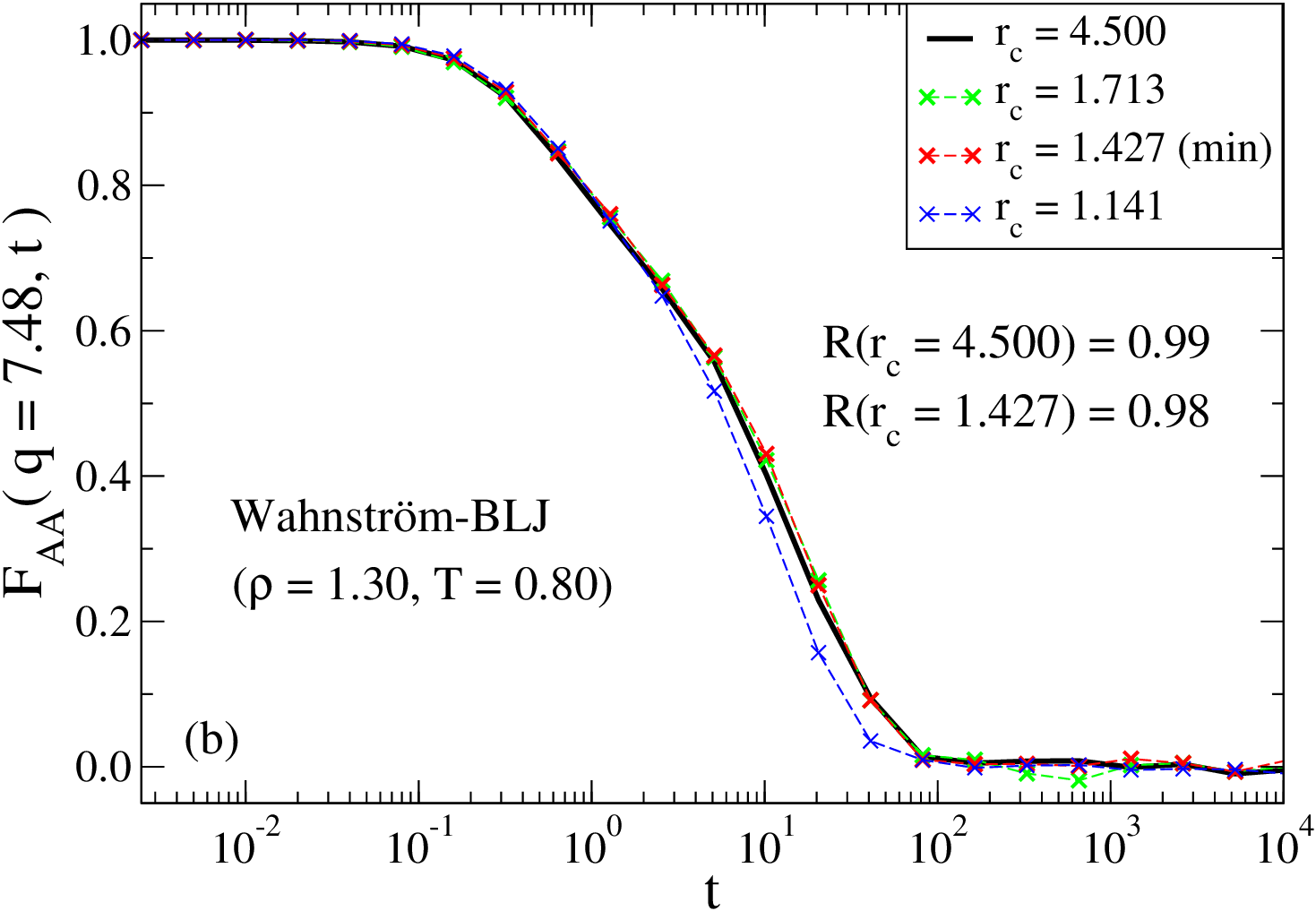}
  \caption{Intermediate coherent scattering function for the Wahnstr{\"o}m 50/50 binary LJ liquid at the wavevector corresponding to the static structure factor maximum. The red and black curves give, respectively, results for a FCS cutoff and large reference cutoff.}
  \label{WaF}
\end{figure}

The role of the cutoff is again investigated by choosing three different cutoffs: one placed at the first minimum of the RDF (red color in Figs. \ref{SCLJgr}-\ref{WaF}), one corresponding to the half height of the RDF from maximum to the first minumum (blue color in Figs. \ref{SCLJgr}-\ref{WaF}), and one displaced to the right of the minimum the same amount as the difference between the first and the second cutoff (green color in Figs. \ref{SCLJgr}-\ref{WaF}). 

In Fig. \ref{SCLJgr} RDFs are shown for the {SCLJ} liquid at three different state points. The red colored circles show results from simulations with an FCS cutoff (marked by the vertical red dashed line), the black curves show the corresponding simulations with a large cutoff (reference system). The insets quantify the deviations in the simulated RDF from the reference  RDF as a function of the cutoff. The reference RDF of Figs. \ref{SCLJgr}(a) and (b) is clearly represented well using an FCS cutoff, while choosing the cutoff inside the FCS results in significant deterioration. At low densities [Figs. \ref{SCLJgr}(c)], deviations occur between FCS cutoff simulations and the reference system. As mentioned, the {SCLJ} liquid is strongly correlating in large parts of its phase diagram, but as density is lowered, the correlations decrease gradually, and the liquid is no longer strongly correlating at state point (c) where $R = 0.50$. These simulations suggest that only when a liquid is strongly correlating is it possible to ignore interactions beyond the FCS.

\begin{figure}[H]
  \centering
  \includegraphics[width=70mm]{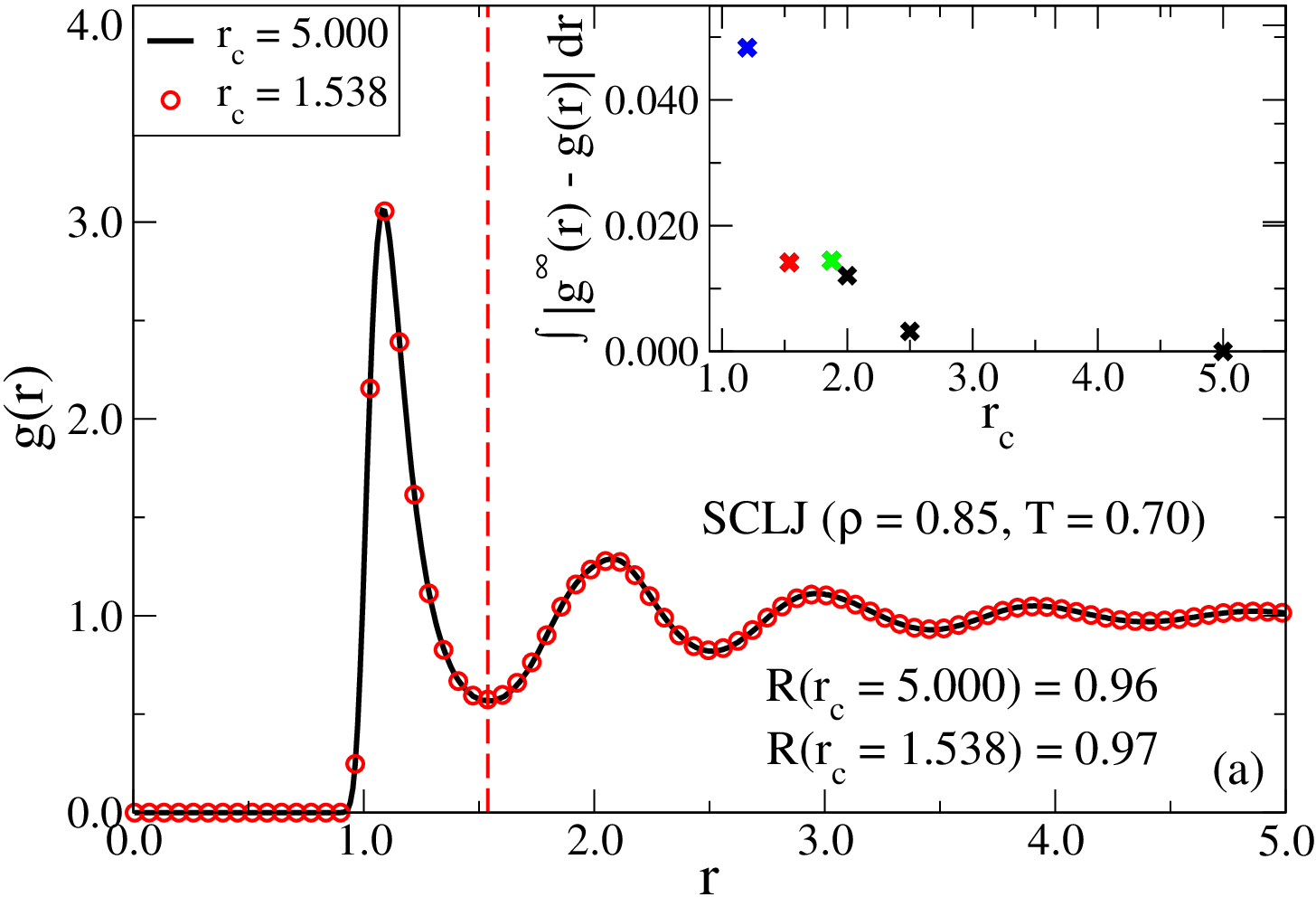}
  \vspace{-20pt}
\end{figure}
\begin{figure}[H]
  \centering
  \includegraphics[width=70mm]{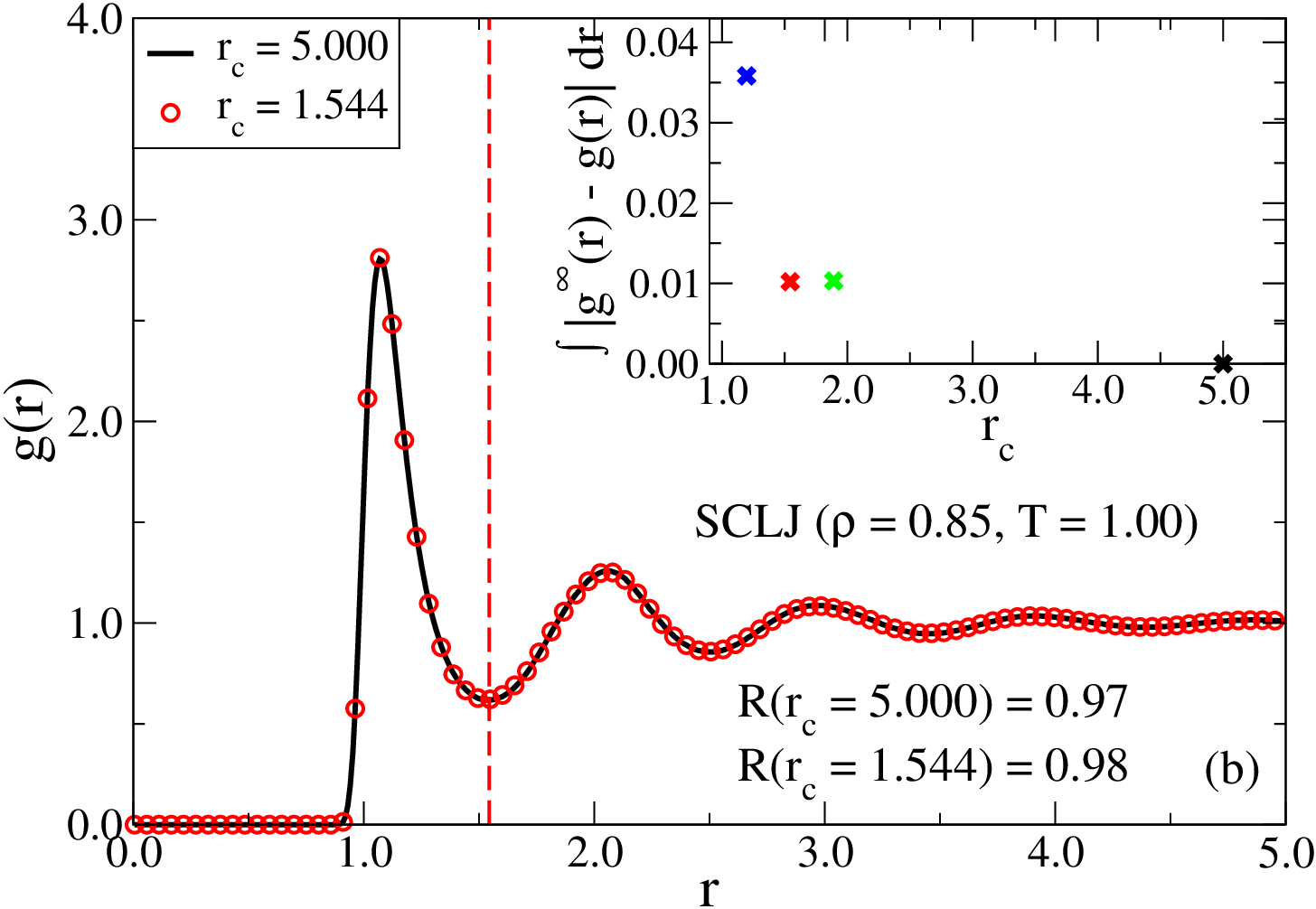}
  \vspace{-20pt}
\end{figure}   

Next, we investigated the SCLJ dynamics at the three state points of Fig. \ref{SCLJgr}. The dynamics is studied via the ISF. The ISFs are shown in Fig. \ref{SCLJdyn}; at all state points the dynamics is represented well using an FCS cutoff.

\begin{figure}[H]
  \centering
  \includegraphics[width=70mm]{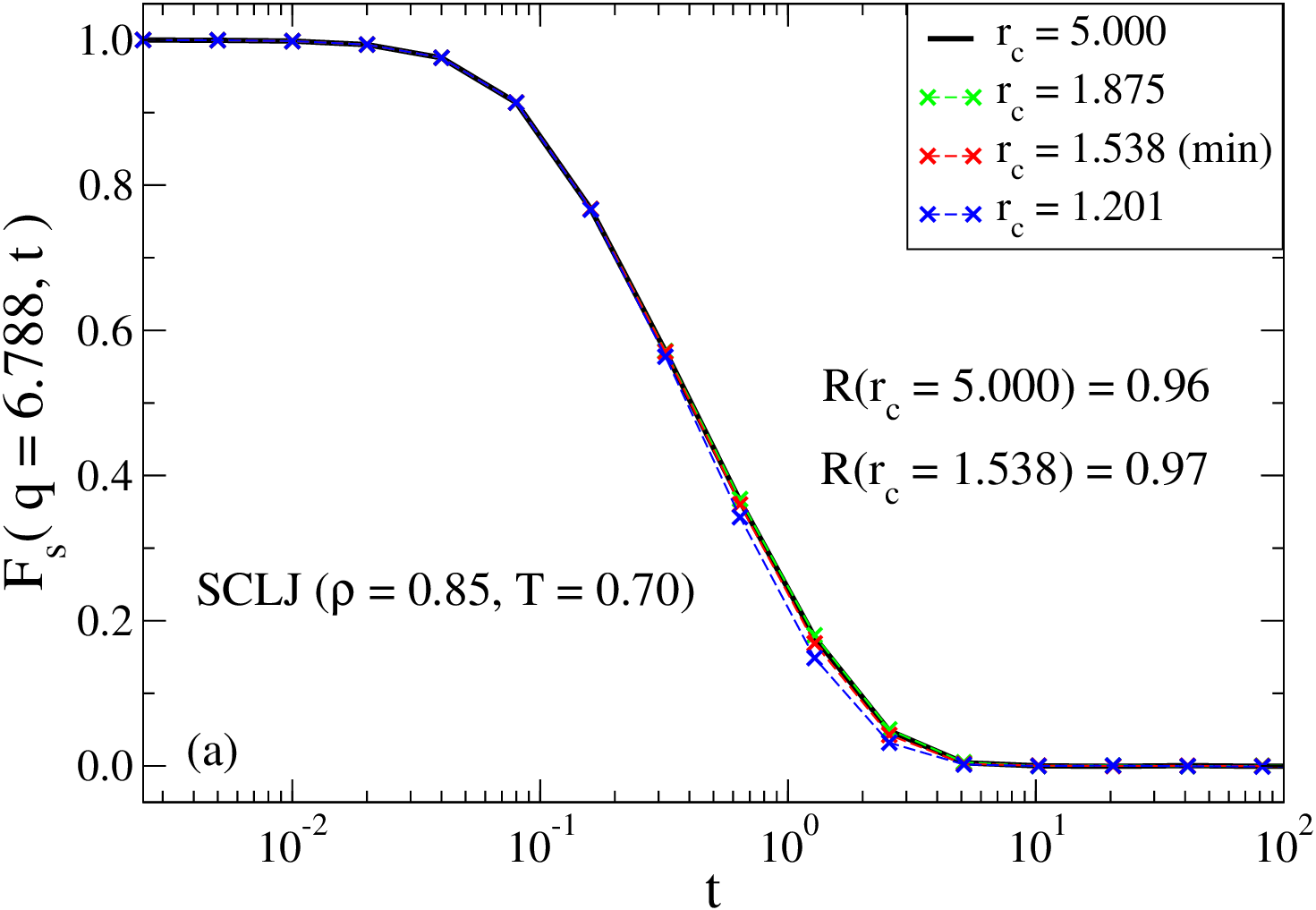}
  \vspace{-20pt}
\end{figure}
\begin{figure}[H]
  \centering
  \includegraphics[width=70mm]{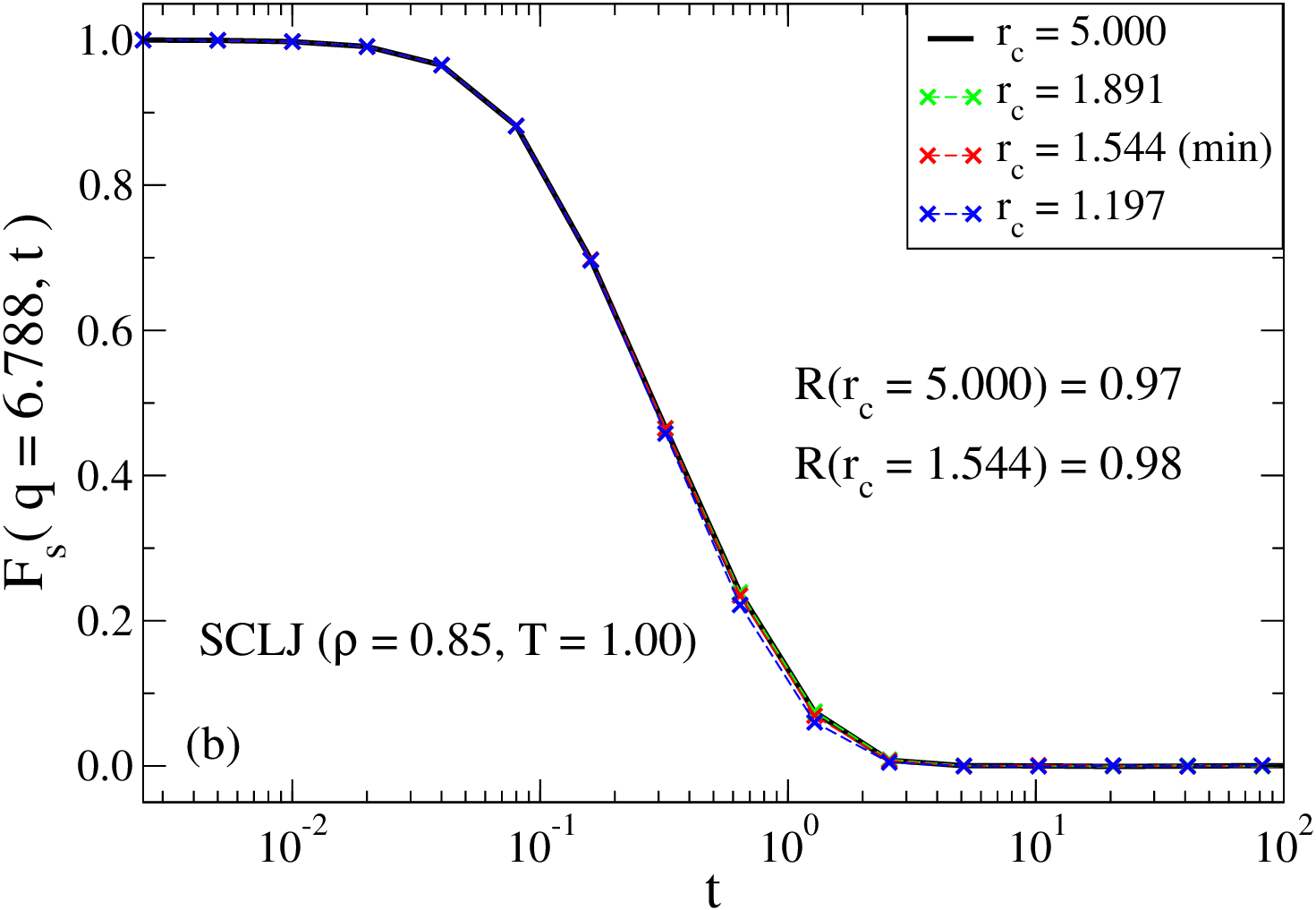}
  \vspace{-20pt}
\end{figure}   

\begin{table}
\begin{tabular}
{|p{1.3cm}||p{0.9cm}|p{0.9cm}|p{0.9cm}||p{2.0cm}|p{2.0cm}|}
\hline
\hline
System & $\rho$ & $T$ & $R$ & $|\Delta RDF|_{max}$ & $|\Delta ISF|_{max}$ \\
\hline
\hline
SCLJ & 0.85 & 1.00 & 0.97 & 1.31$\cdot10^{-2}$ & 5.10$\cdot10^{-3}$ \\
SCLJ & 0.85 & 0.70 & 0.96 & 1.68$\cdot10^{-2}$ & 8.28$\cdot10^{-3}$ \\
SCLJ & 0.85 & 0.65 & 0.96 & 1.63$\cdot10^{-2}$ & 8.96$\cdot10^{-3}$ \\
SCLJ & 0.50 & 1.50 & 0.69 & 11.2$\cdot10^{-2}$ & 7.94$\cdot10^{-3}$ \\
SCLJ & 0.55 & 1.13 & 0.50 & 15.2$\cdot10^{-2}$ & 12.0$\cdot10^{-3}$ \\
\hline
\hline
\end{tabular}
\caption{Simulation results for five state points of the SCLJ liquid. For each state point is given density, temperature, correlation coefficient, maximum deviation from the true RDF using an FCS cutoff, and maximum deviation from the true incoherent structure factor using an FCS cutoff.}
\label{table}
\end{table}

Table \ref{table} summarizes results for five state points of the SCLJ liquid. The third state point is slightly below the triple point, i.e., in a slightly metastable state. The deviations clearly increase as the $WU$ correlations decrease.

We proceed to investigate mixtures of two different particles (A and B) interacting with LJ type potentials. The cutoff used for all three interactions (AA, AB, BB) are placed at the same distance, referring to $\sigma_{AA}$. In Fig. \ref{KABLJgr} reference and FCS cutoff results are shown for the $AA$-particle RDFs of generalized KABLJ mixtures with repulsive exponent $12$ and attractive exponents $n=4,10$. The $n=4$ case is below our definition of a strongly correlating liquid $R>0.9$ (at the simulated state point this system has negative virial), but is still pretty well correlating. For all investigated state points an FCS cutoff gives accurate results. We found the same using the standard repulsive exponent $n=6$ \cite{FCS2} (results not shown).

\begin{figure}[H]
  \centering
  \includegraphics[width=70mm]{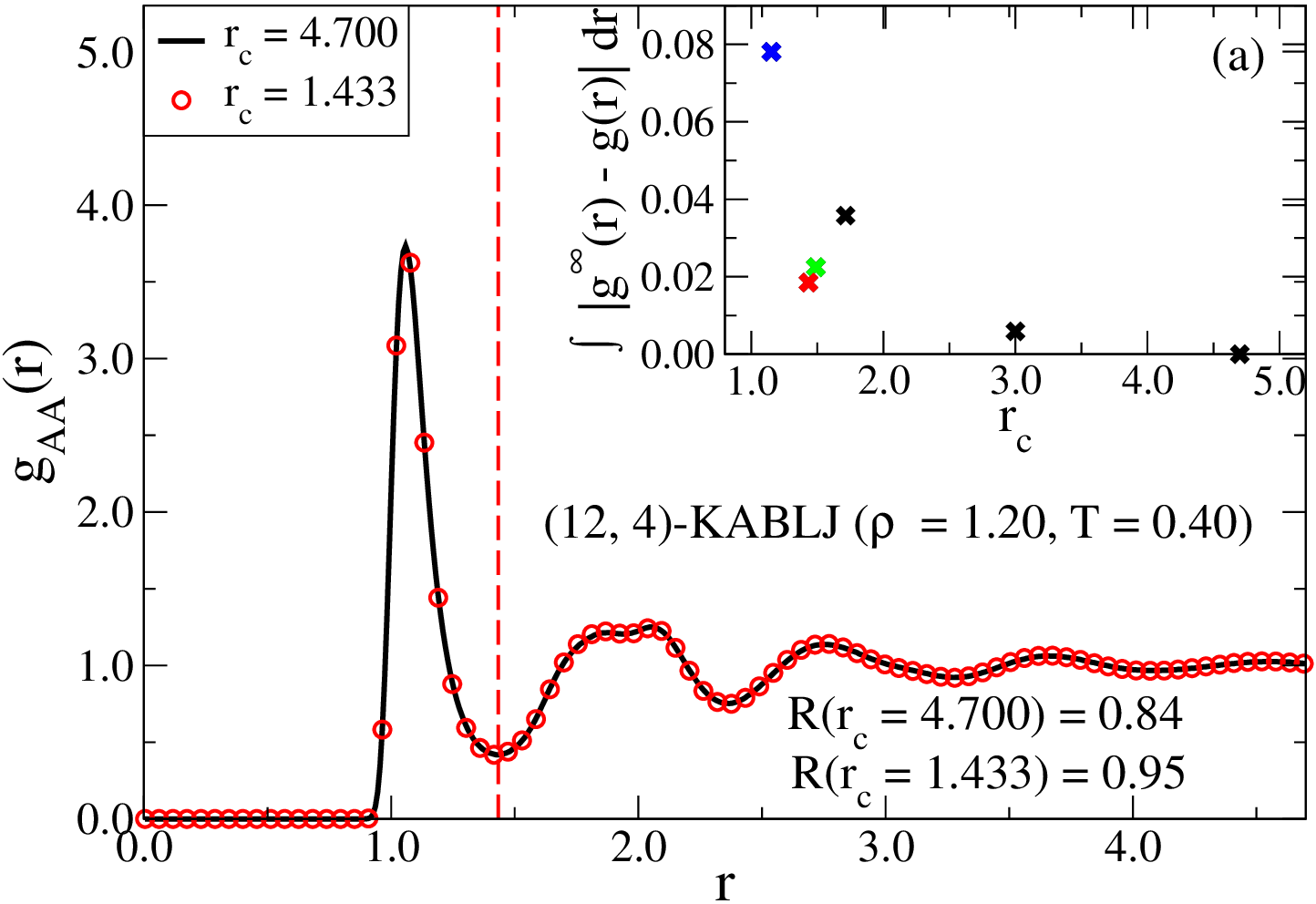}
  \vspace{-20pt}
\end{figure}   

The A-particle {ISF}s for the state points of Fig. \ref{KABLJgr} are shown in Fig. \ref{KABLJFsA}. For the KABLJ mixture,  placing the cutoff inside the FCS (blue curves) fails to reproduce the dynamics properly, whereas it is well approximated using an FCS cutoff (red). Slight deviations are noted for the red curves, an issue considered in Appendix \ref{FCS_sec} that discusses alternatives for delimiting the FCS. Similar results are found for the B particles (results not shown). 

\begin{figure}[H]
  \centering
  \includegraphics[width=70mm]{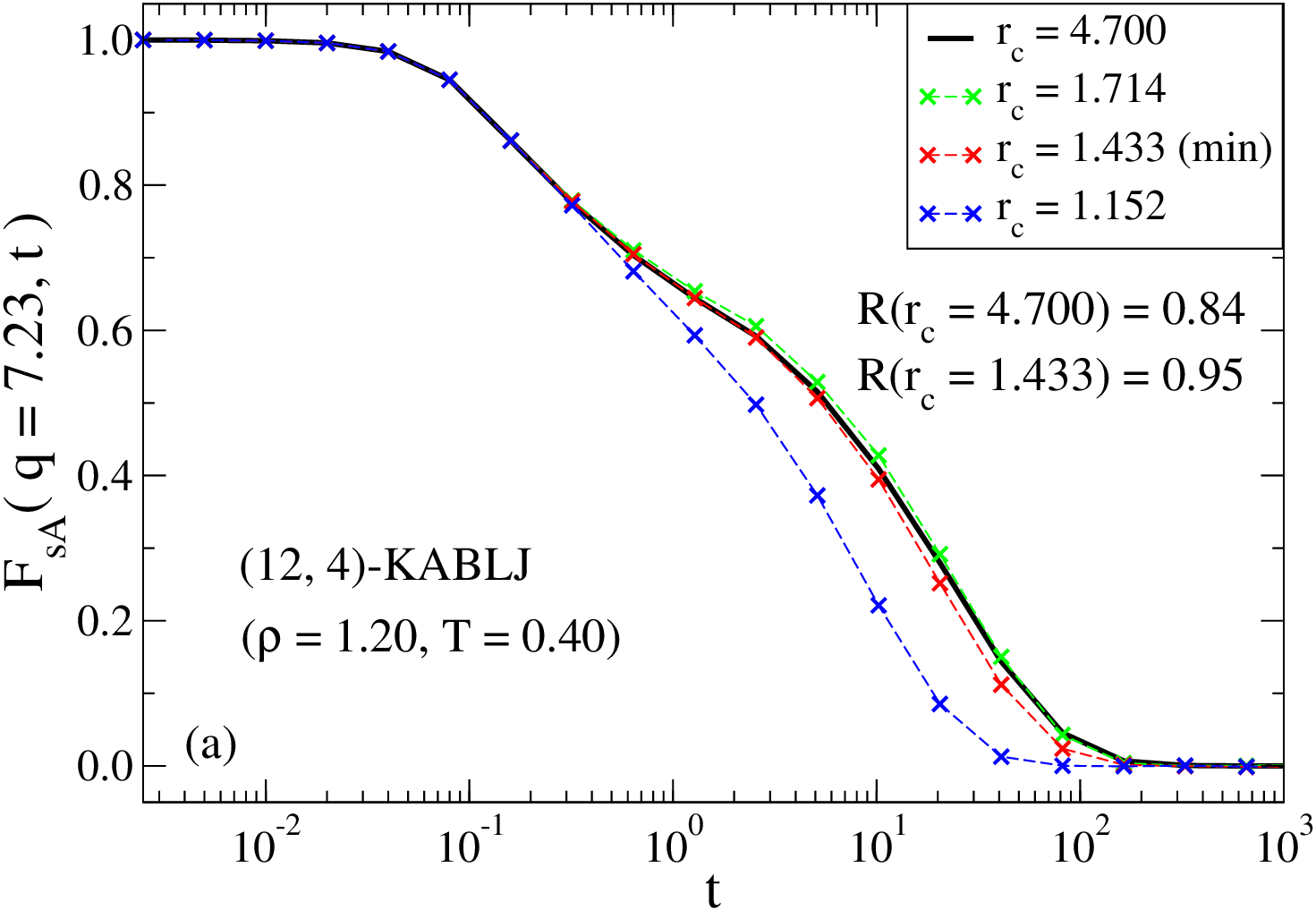}
  \vspace{-20pt}
\end{figure}    

We also simulated the Wahnstr{\"o}m 50-50 binary LJ mixture \cite{Wahnstromblj}, finding again that whenever $R>0.9$ structure and dynamics are well reproduced using an FCS cutoff. We do not show these results, but show instead results for the coherent scattering function at one state point (Fig. \ref{WaF}). Again, the FCS cutoff (red crosses) gives the correct dynamics whereas reducing the cutoff further does not give proper results (blue crosses).

In summary, for all LJ type systems whenever there are strong virial potential-energy correlations ($R>0.9$), an FCS cutoff gives accurate results for both structure and dynamics. In the parts of the phase diagram where $WU$ correlations are weaker, an FCS cutoff gives poor results.

\subsection{Buckingham liquid}

\begin{figure}[H]
  \centering
  \includegraphics[width=70mm]{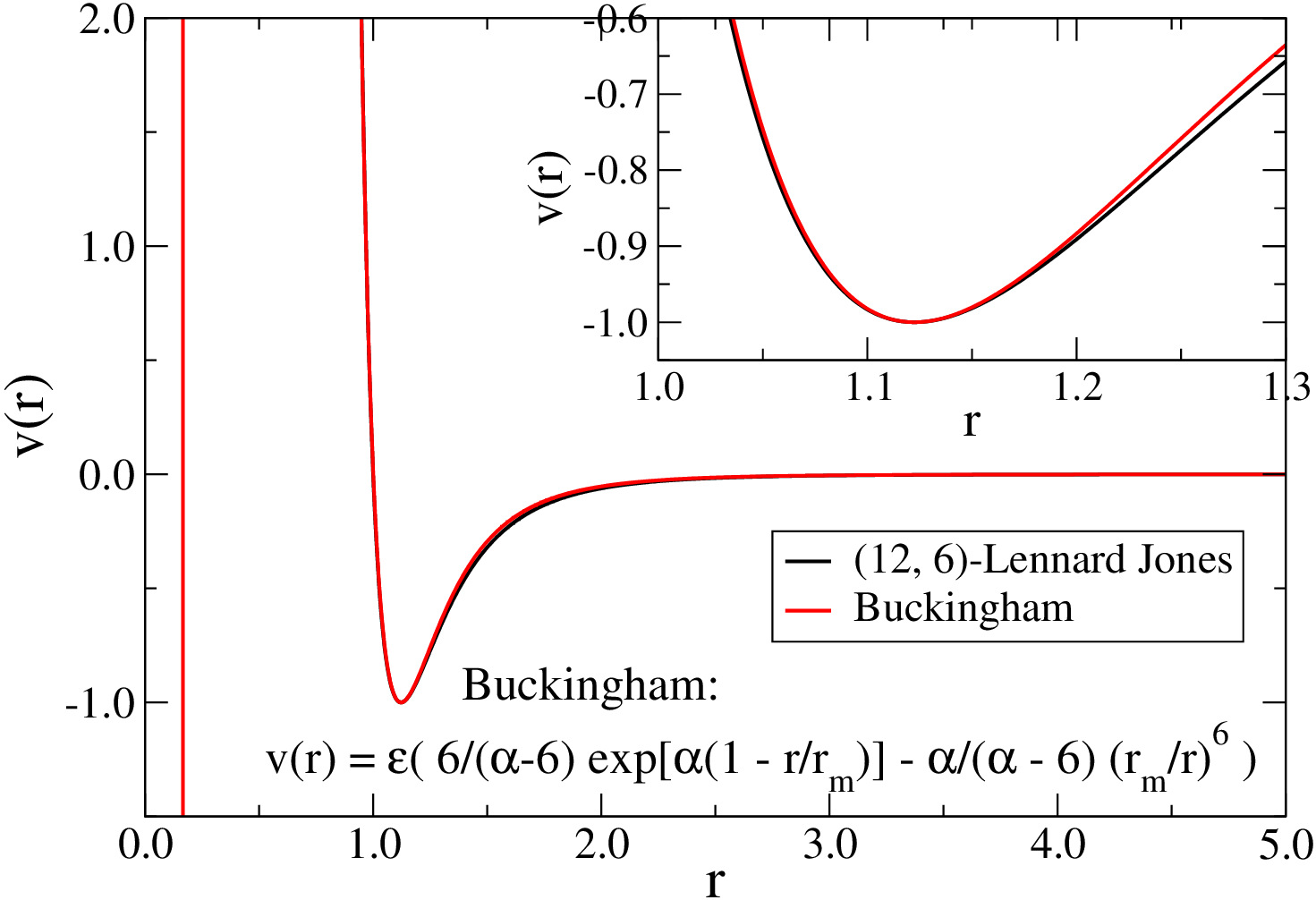}
  \caption{The Buckingham pair potential (red) and the LJ pair potential (black). The parameters of the Buckingham potential were chosen such that the LJ potential is well approximated in the repulsive region.}
  \label{pairSCB}
\end{figure}

Next, we consider the single-component Buckingham liquid ({SCB}). The Buckingham potential \cite{buckingham1,buckinghamisomorphs} is similar to the LJ potential, but does not have an IPL repulsive term; instead the potential's short-distance behavior follows a steep exponential (Fig. \ref{pairSCB}). The parameters of the Buckingham potential (Appendix \ref{pairP}) were chosen such that the LJ potential is well approximated in the repulsive region (Fig. \ref{pairSCB}).

\begin{figure}[H]
  \centering
  \includegraphics[width=70mm]{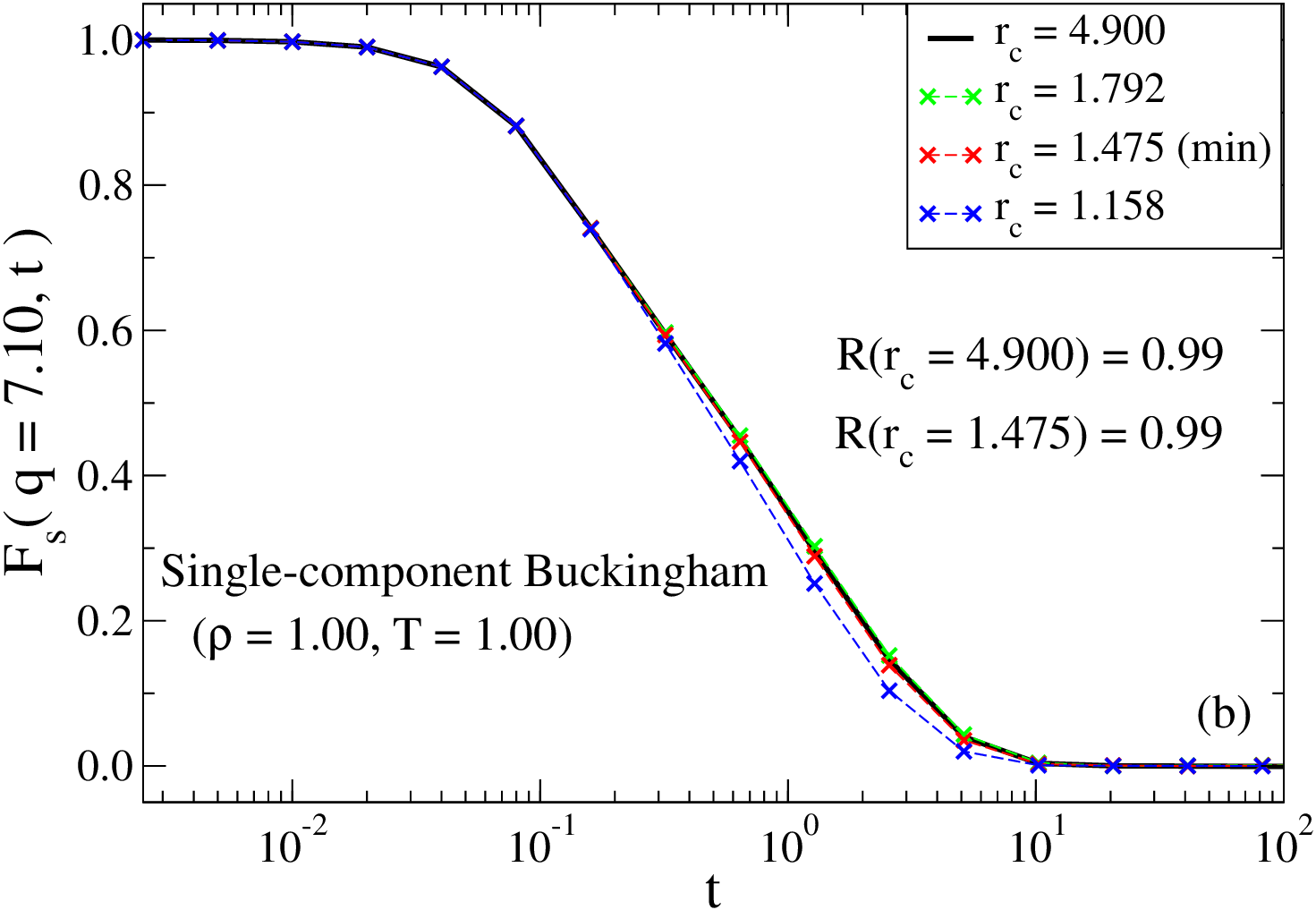}
  \caption{The effect on structure and dynamics of varying the cutoff for the single-component Buckingham (SCB) liquid. The red and black curves give, respectively, results for an FCS cutoff and a large reference cutoff.
    (a) RDF at $\rho=1.00$ and $T = 1.00$ ($R=0.99$). The inset quantifies the deviation in RDF from the reference RDF as a function of the cutoff. 
  (b) {ISF} at the same state point.}
  \label{SCB}
\end{figure}

Figures \ref{SCB}(a) and (b) show, respectively, the RDF and ISF for the SCB liquid. The FCS cutoff works well.

\begin{figure}[H]
  \centering
  \includegraphics[width=70mm]{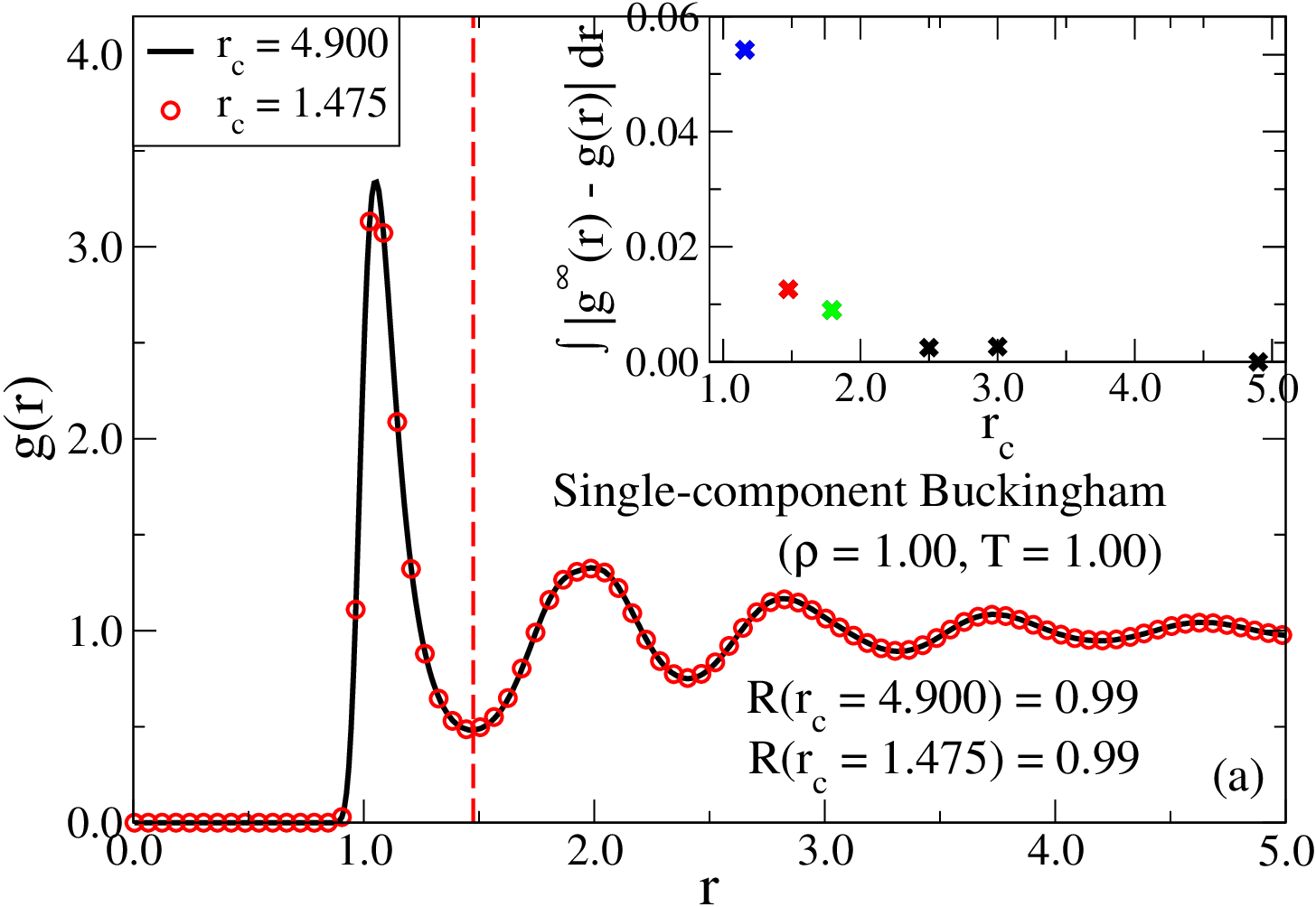}
  \vspace{-20pt}
\end{figure}

\subsection{Dzugutov liquid}

\begin{figure}[H]
  \centering
  \includegraphics[width=70mm]{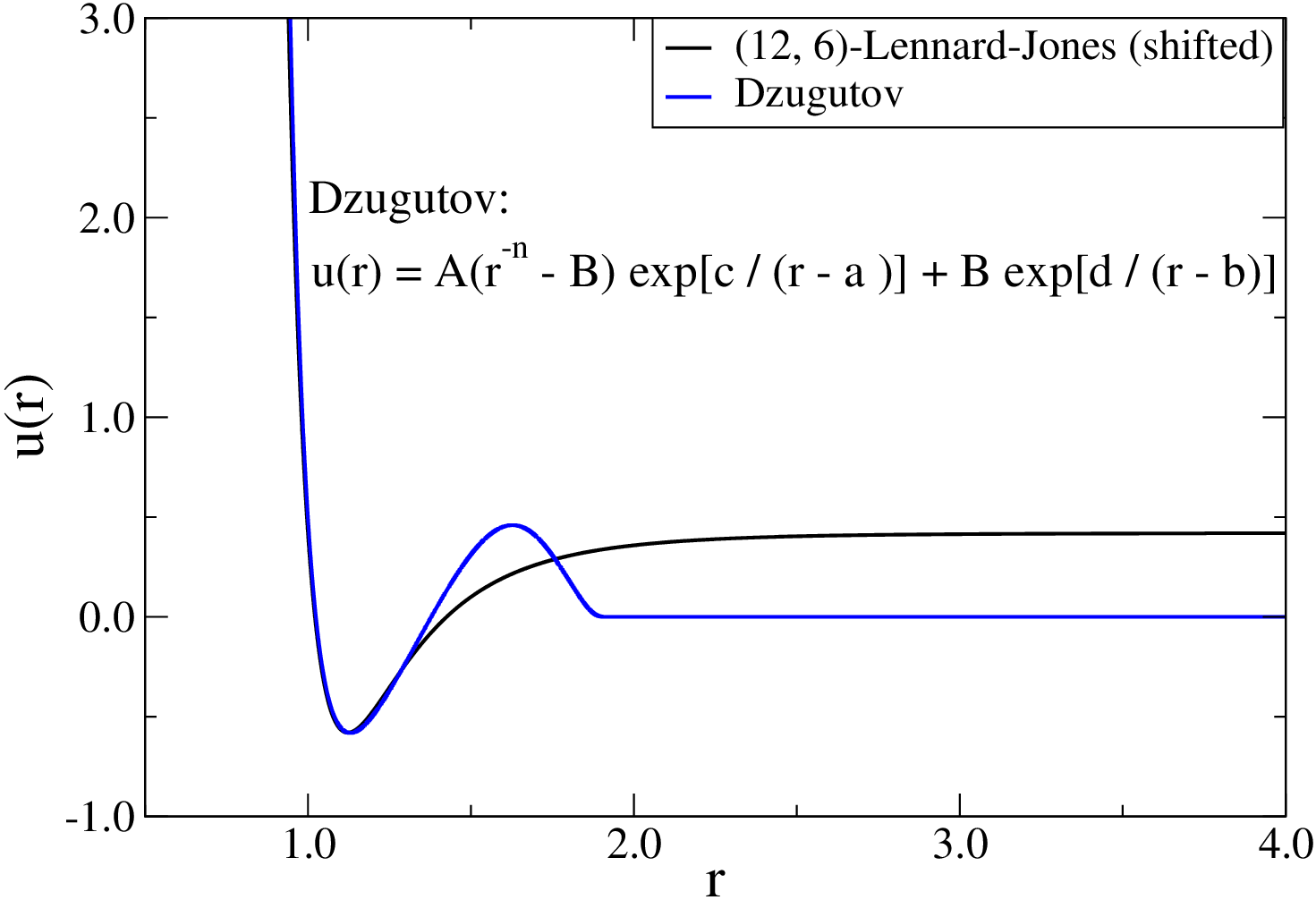}
  \caption{The Dzugutov ({DZ}) pair potential \cite{dzugutov} (blue). Also shown is the LJ pair potential (black curve). The {DZ} potential approximates the LJ potential around the first minimum, but has a local maximum at larger distances.}
  \label{pairDZ}
\end{figure}

Figure \ref{pairDZ} shows the Dzugutov (DZ) pair potential \cite{dzugutov}, which was originally suggested as a model potential impeding crystallization by energetically punishing particle separations corresponding to the next-nearest neighbor distance of a face-centered cubic lattice. At short distances the {DZ} pair potential approximates the LJ potential.

\begin{figure}[H]
  \centering
  \includegraphics[width=70mm]{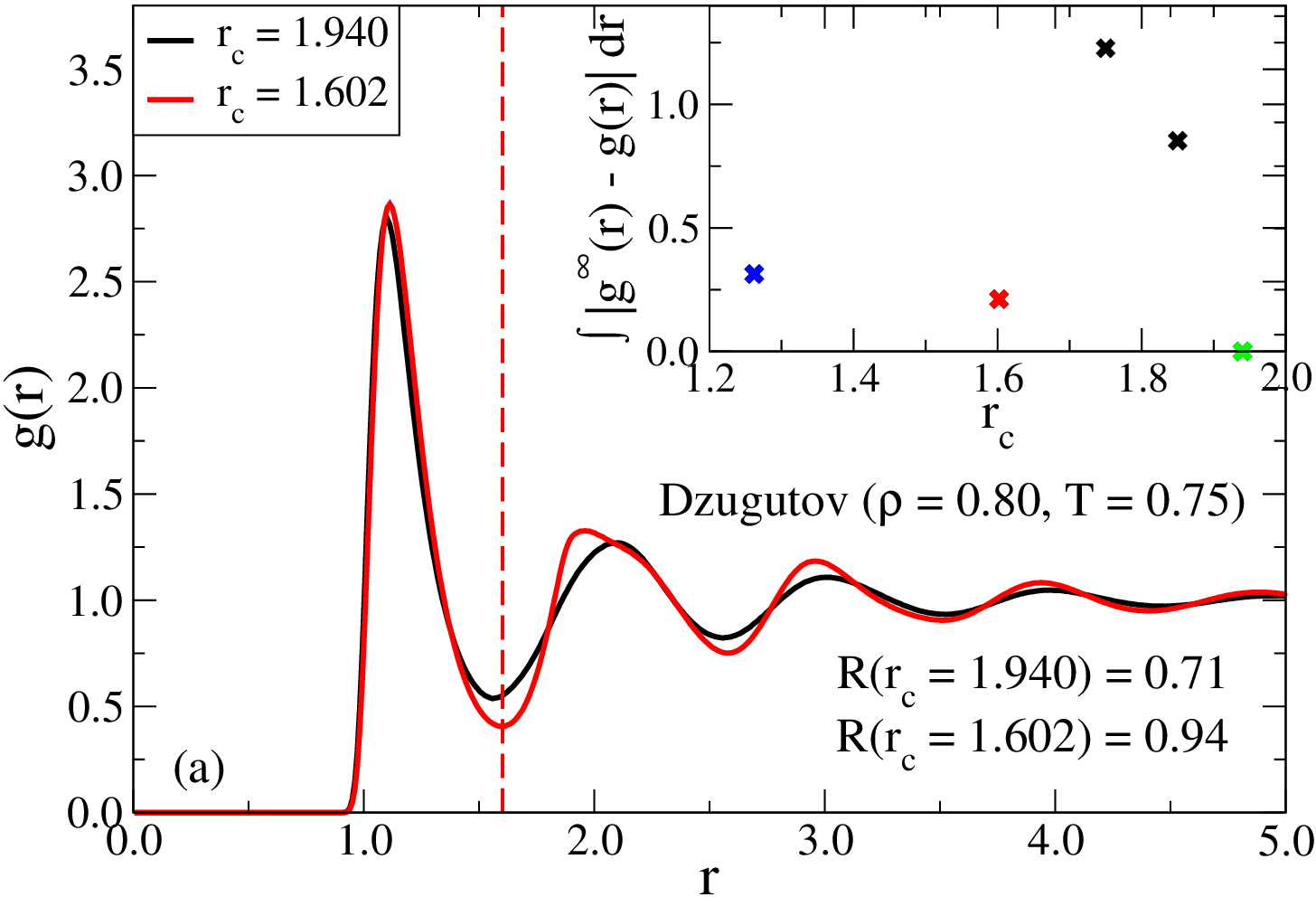}
  \includegraphics[width=70mm]{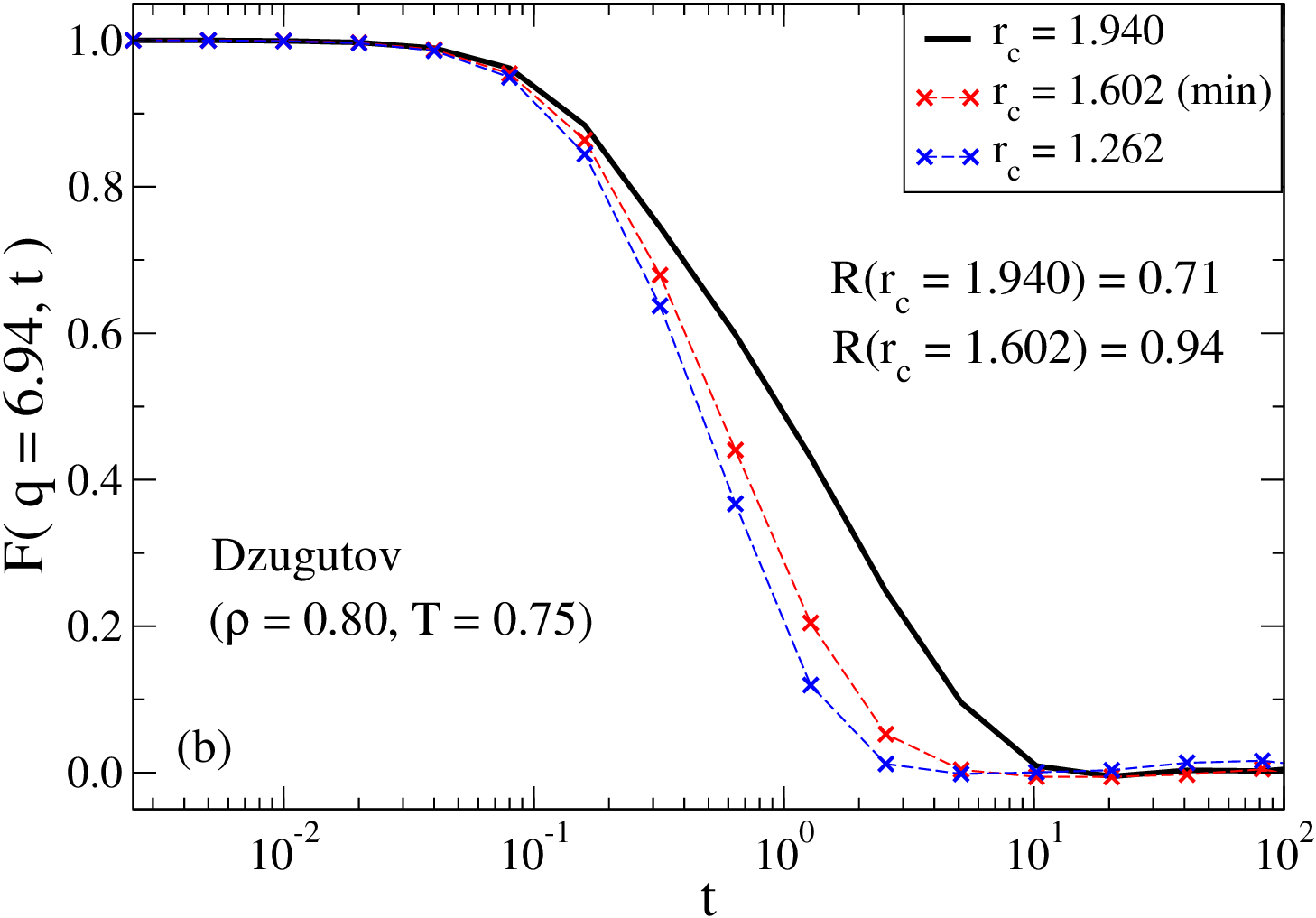}
  \vspace{0.5cm}
  \caption{The effect on structure and dynamics of varying the cutoff for the Dzugutov (DZ) liquid at $\rho=0.80$ and $T = 0.75$ ($R=0.71$). The red and black curves give, respectively, results for an FCS cutoff and a large reference cutoff.
    (a) RDF where the inset quantifies the deviation from the reference RDF as a function of the cutoff. 
(b) The intermediate coherent scattering function at the same state point, including here results for a cutoff within the FCS (blue crosses).}
  \label{DZ}
\end{figure}

Figures \ref{DZ}(a) and (b) show, respectively, the RDF and intermediate coherent scattering function of the DZ system. For this system the use of an FCS cutoff leads to poor results. This is not surprising given the fact that using an FCS cutoff removes the maximum of the {DZ} potential. What is important here, however, is that the poor FCS cutoff results correlate with the fairly weak virial/potential-energy correlations ($R=0.71$). 

\subsection{Lennard-Jones Gaussian liquid}

\begin{figure}[H]
  \centering
  \includegraphics[width=70mm]{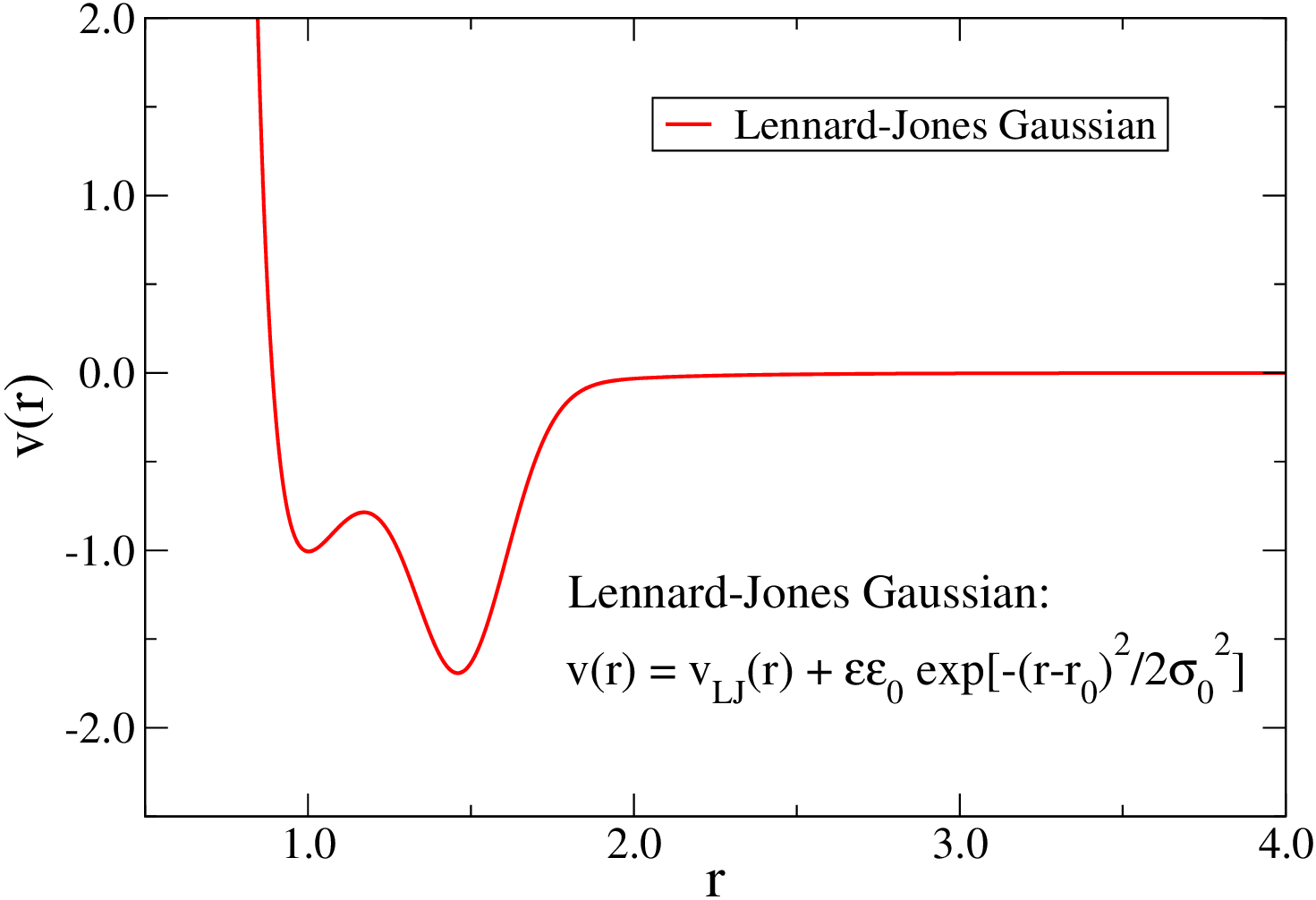}
  \caption{The Lennard-Jones Gaussian (LJG) pair potential \cite{LJG} constructed by adding a Gaussian to the pair LJ potential. Two distinct minima are present.}
  \label{pairLJG}
\end{figure}

The Lennard-Jones Gaussian liquid \cite{LJG} (LJG) is a non-strongly correlating liquid with the two-minimum pair potential  shown in Fig. \ref{pairLJG}. The parameters of LJG model (Appendix \ref{pairP}) are such that the second LJG potential minimum does not coincide with that of the {SCLJ} system \cite{ljgglassformer}.

\begin{figure}[H]
  \centering
  \includegraphics[width=70mm]{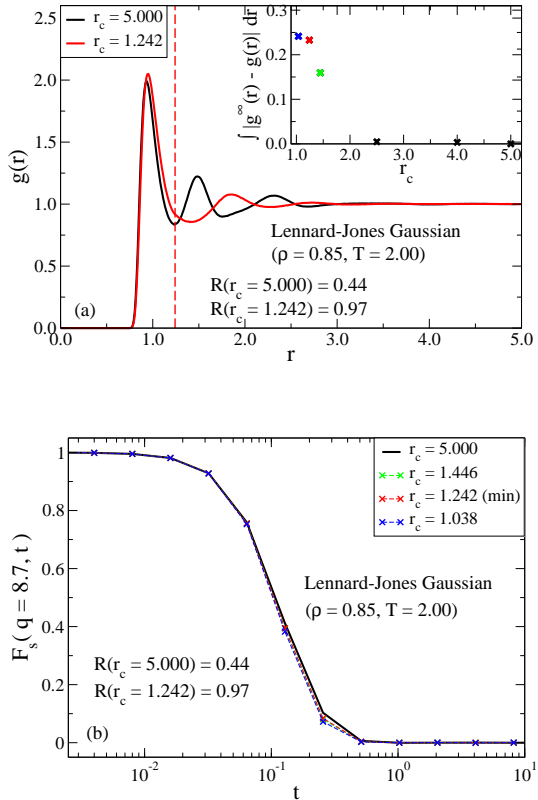}
  \caption{The effect on structure and dynamics of the cutoff for the LJG liquid.
    The red and black curves give, respectively, results for an FCS cutoff and a large reference cutoff.
    (a) RDF at $\rho=0.85$ and $T = 2.00$ ($R=0.44$). The inset quantifies the deviation in RDF from the reference RDF (black curve) as a function of the cutoff. 
(b) {ISF} at the same state point.}
  \label{LJG}
\end{figure}

Results from simulating structure and dynamics of the LJG liquid are shown in Figs. \ref{LJG}(a) and (b). The  FCS cutoff does not give the correct RDF, whereas deviations in the dynamics are fairly small. Note that the FCS cutoff removes the second minimum. 

\begin{figure}[H]
  \centering
  \includegraphics[width=70mm]{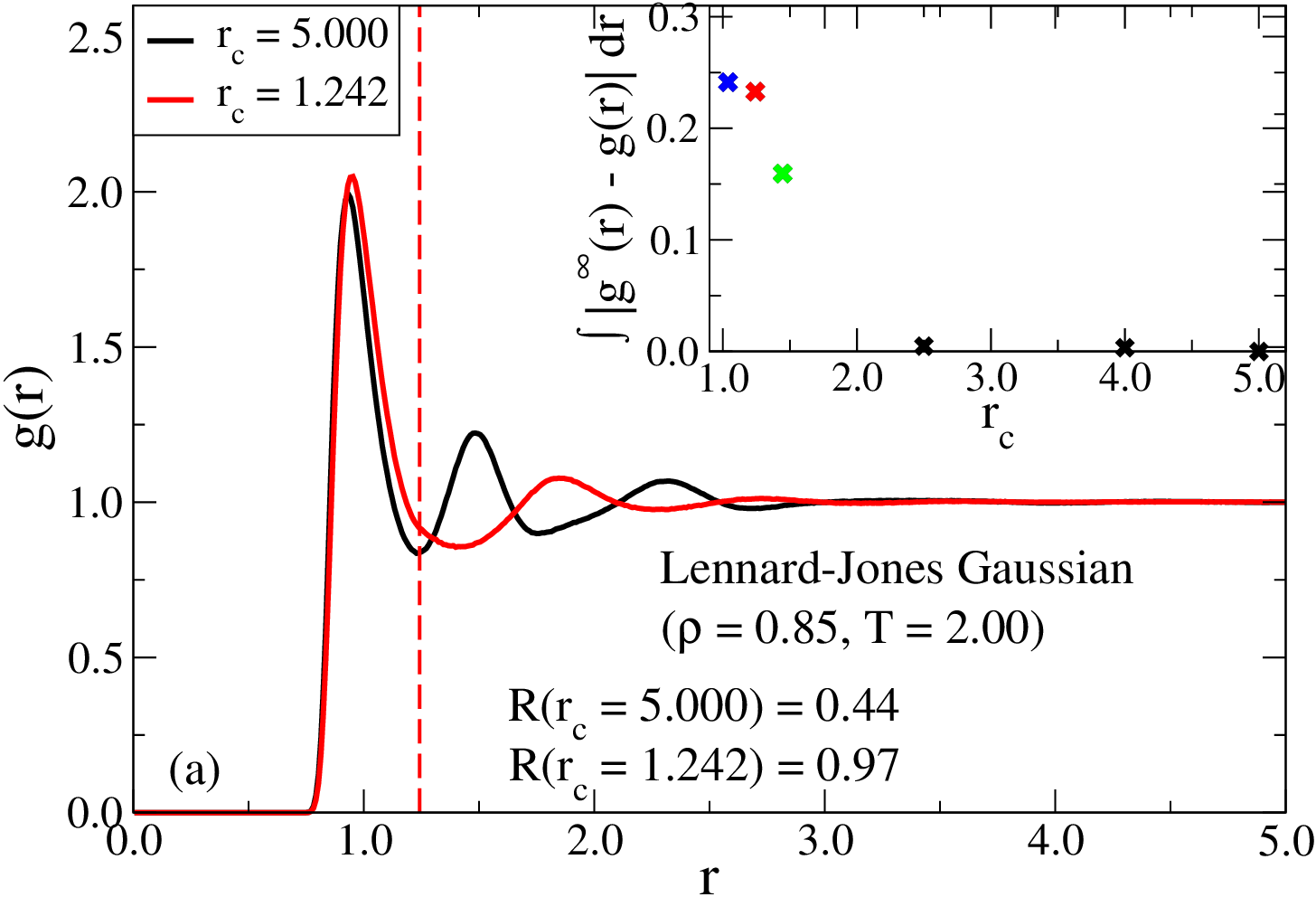}
  \vspace{-20pt}
\end{figure}

\subsection{Gaussian core model}

The Gaussian core model (GCM) \cite{gcm1,gcm2}, which is not strongly correlating, is defined by a Gaussian pair potential and thus has a finite potential energy at zero separation. The high-density regime of the {GCM} model ($\rho > 1.5$) has recently received attention as a single-component model glass former \cite{highdensitygcm} because it is not prone to crystallization and shows the characteristic features of glass-forming liquids (large viscosity, two-step relaxation, etc).

\begin{figure}[H]
  \centering
  \includegraphics[width=70mm]{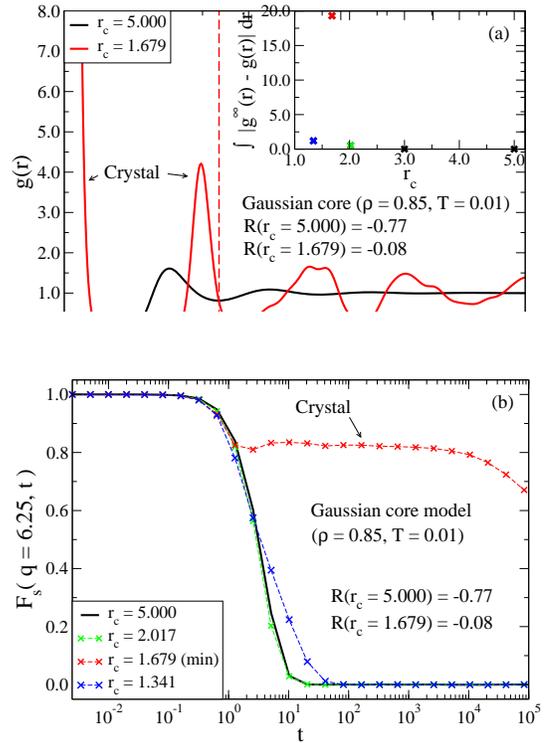}
  \caption{The effect on structure and dynamics of varying the cutoff for the GCM liquid.  The red and black curves give, respectively, results for an FCS cutoff and a large reference cutoff.
    (a) RDF at $\rho=0.85$ and $T = 0.01$ ($R=-0.77$). The inset quantifies the deviation in RDF from the reference RDF as a function of the cutoff. The red curve represents a crystallized state.
(b) {ISF} at the same state points.}
  \label{GCM}
\end{figure}

Figure \ref{GCM} shows the RDF and {ISF} for the {GCM} liquid. The {GCM} crystallizes when an FCS cutoff is used. For this reason, obviously, an FCS cutoff is not able to reproduce structure and dynamics of the reference system. 

\begin{figure}[H]
  \centering
  \includegraphics[width=70mm]{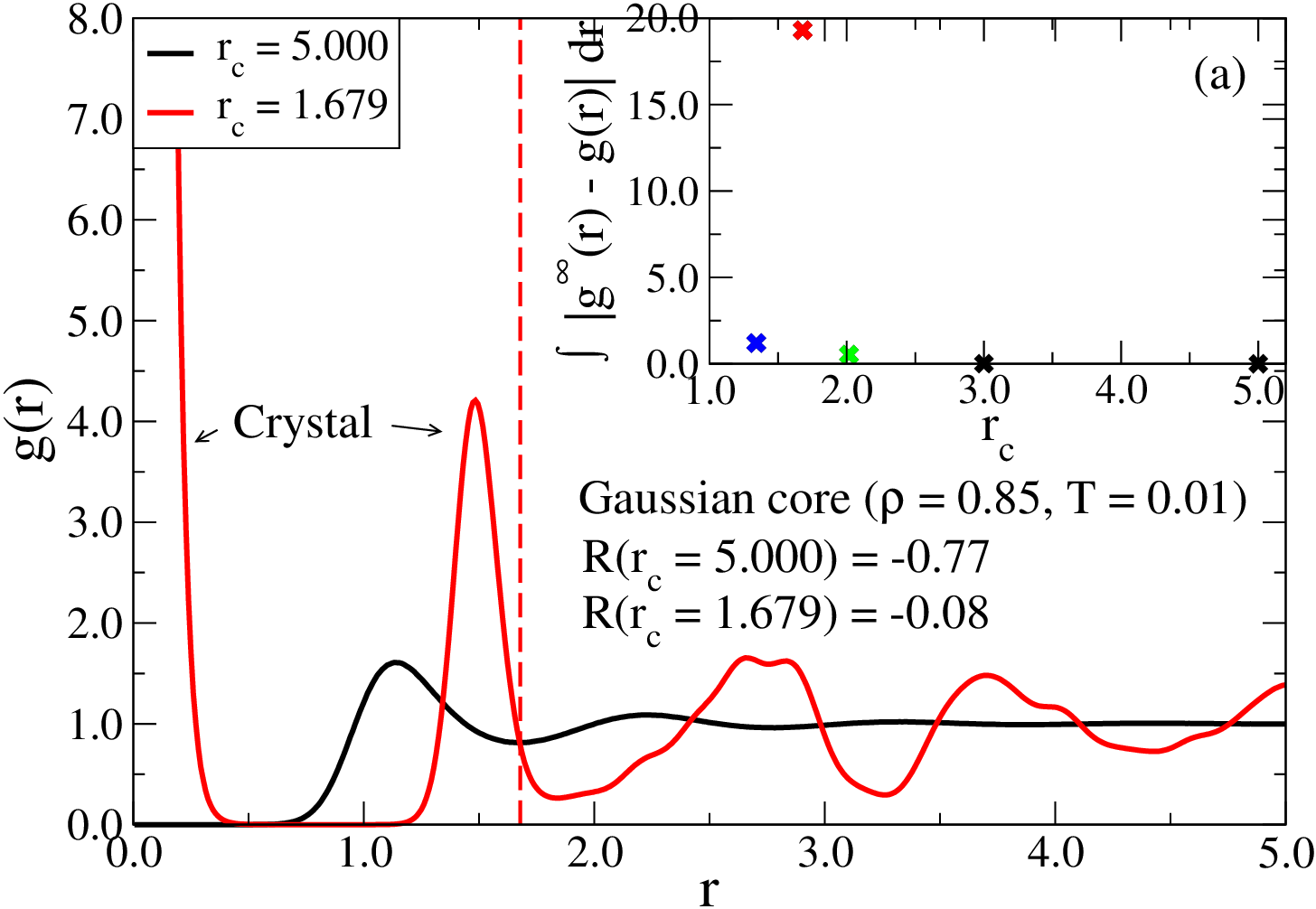}
  \vspace{-20pt}
\end{figure}

\subsection{The Hansen-McDonald molten salt model}

\begin{figure}[H]
  \centering
  \includegraphics[width=70mm]{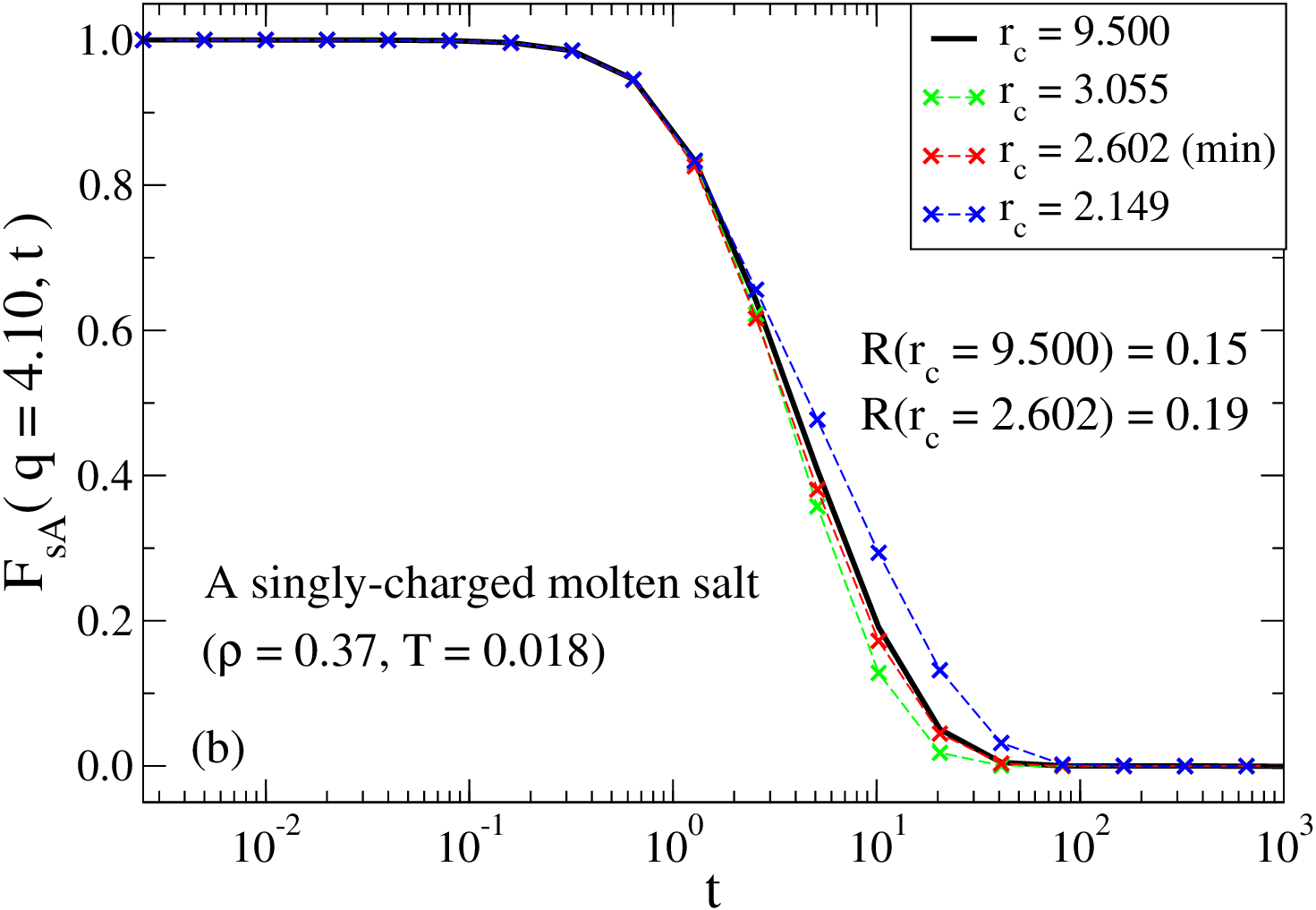}
 \caption{The effect on structure and dynamics of varying the cutoff for Hansen-McDonald singly-charged molten salt model. Because of the competing interactions (Coulomb and $n=9$ repulsive IPL) this model is not strongly correlating.  The red and black curves give, respectively, results for an FCS cutoff and a large reference cutoff.
    (a) $AA$-particle RDF at $\rho=0.37$ and $T = 0.018$ ($R=0.15$). The inset quantifies the deviation in RDF from the reference RDF as a function of the cutoff. 
(b) $A$-particle {ISF} at the same state point.}
  \label{MS}
\end{figure}

The final atomic system studied is the ``singly-charged molten salt model'' of Hansen and McDonald \cite{moltensalt}. In Fig. \ref{MS} we see that the structure is not represented well by the use of an FCS cutoff. Interestingly, the dynamics {\it is} well reproduced using this cutoff -- even better, in fact, than for a larger cutoff (Fig. \ref{MS}(b)).

\begin{figure}[H]
  \centering
  \includegraphics[width=70mm]{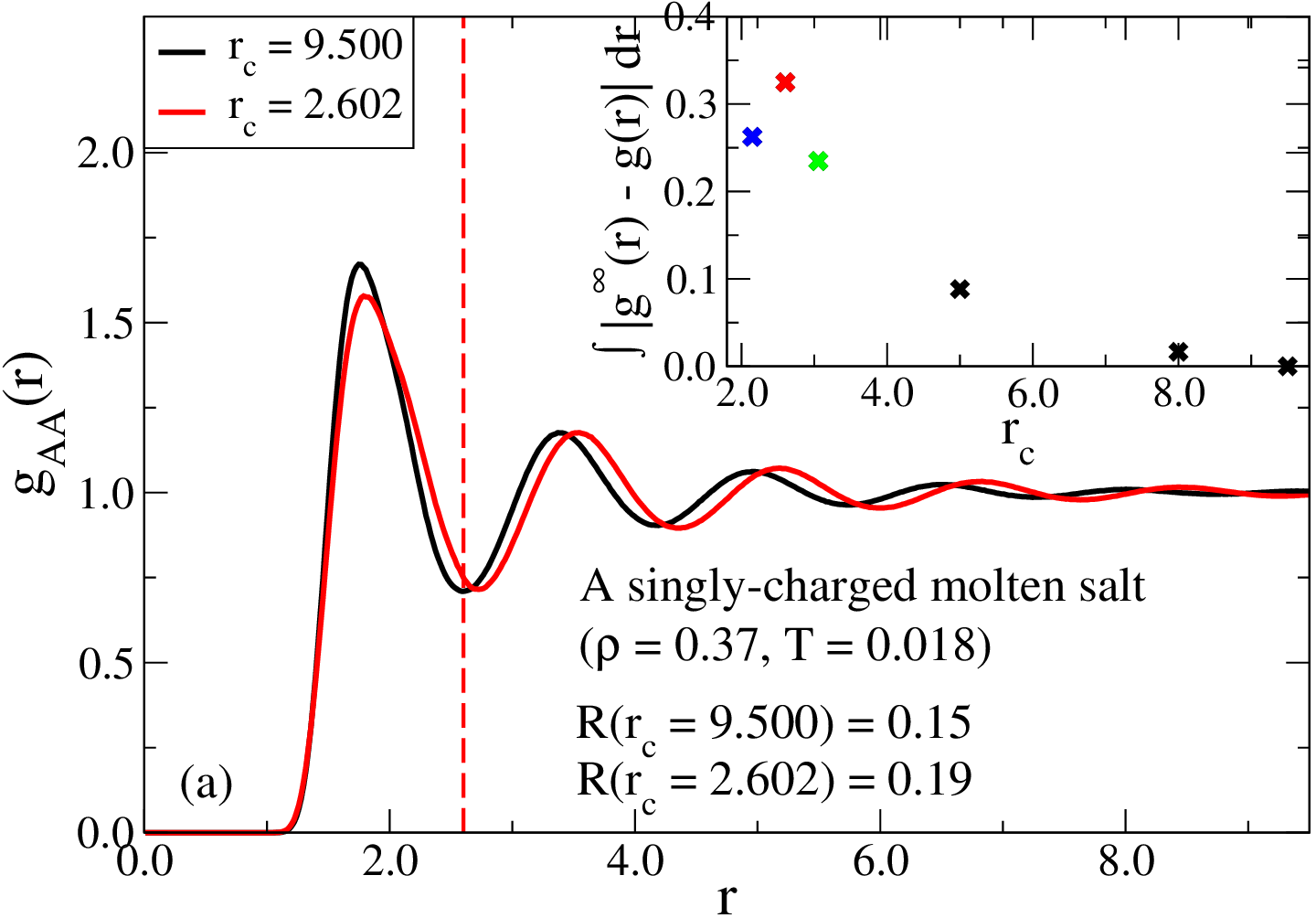}
  \vspace{-20pt}
\end{figure}

\subsection{Two strongly correlating molecular model liquids}

We finish the presentation of the numerical results by giving data for three molecular model liquids. In this subsection data are given for two strongly correlating molecular liquid models, the Lewis-Wahnstr{\"o}m OTP \cite{otp1,otp2} and the asymmetric dumbbell \cite{sch09} models, which represent a molecule by three and two rigidly bonded LJ spheres, respectively. The next subsection gives data for a rigid water model. 

\begin{figure}[H]
  \centering
  \includegraphics[width=70mm]{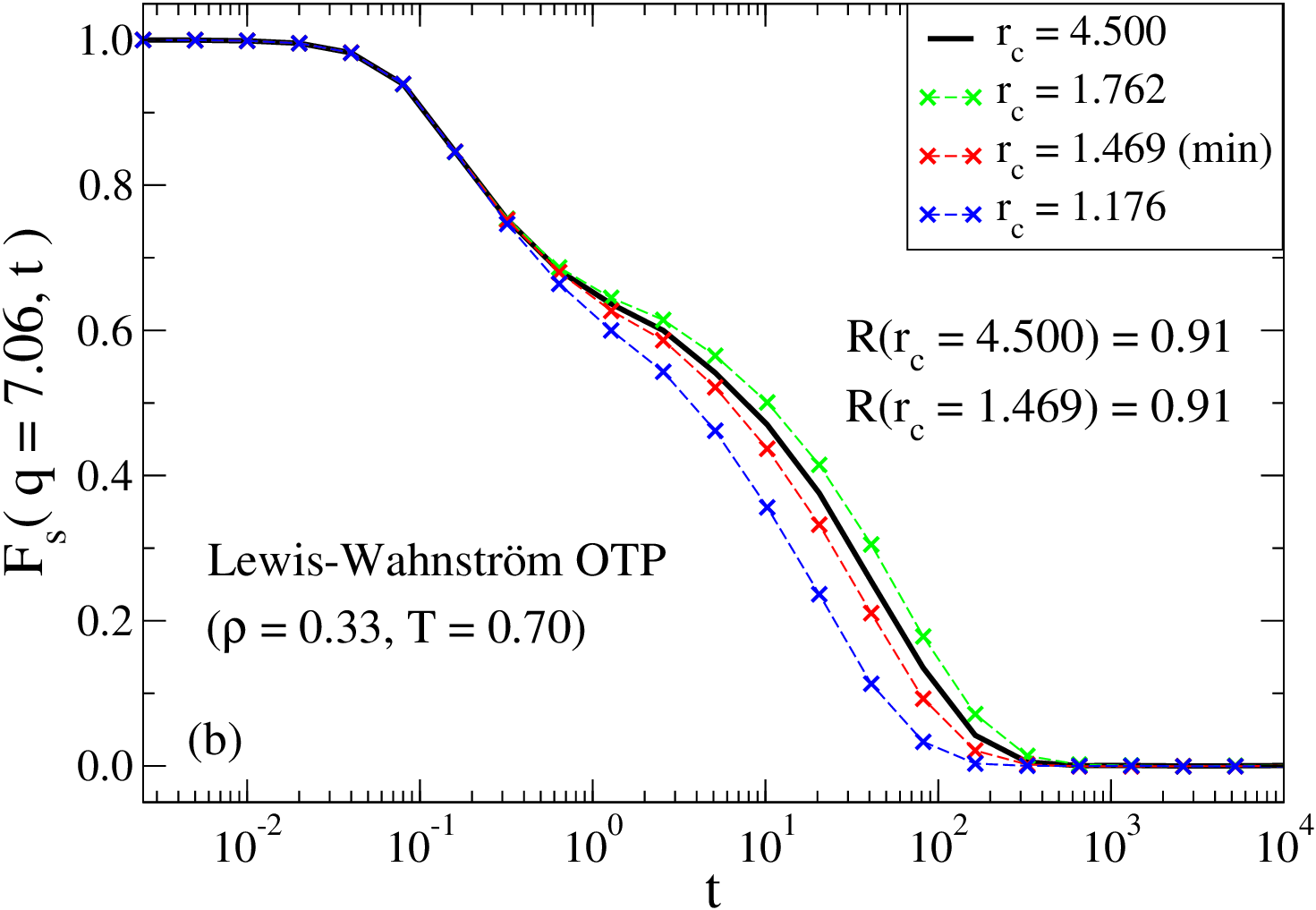}
  \caption{The effect on structure and dynamics of varying the cutoff for the Wahnstr{\"o}m OTP model. The red and black curves give, respectively, results for an FCS cutoff and a large reference cutoff.
(a) RDF of the LJ particles at $\rho=0.33$ and $T = 0.70$ ($R=0.91$). The inset quantifies the deviation in RDF from the reference RDF  as a function of the cutoff. The spikes derive from the bonds.
(b) {ISF} at the same state point.}
  \label{OTP}
\end{figure}

\begin{figure}[H]
  \centering
  \includegraphics[width=70mm]{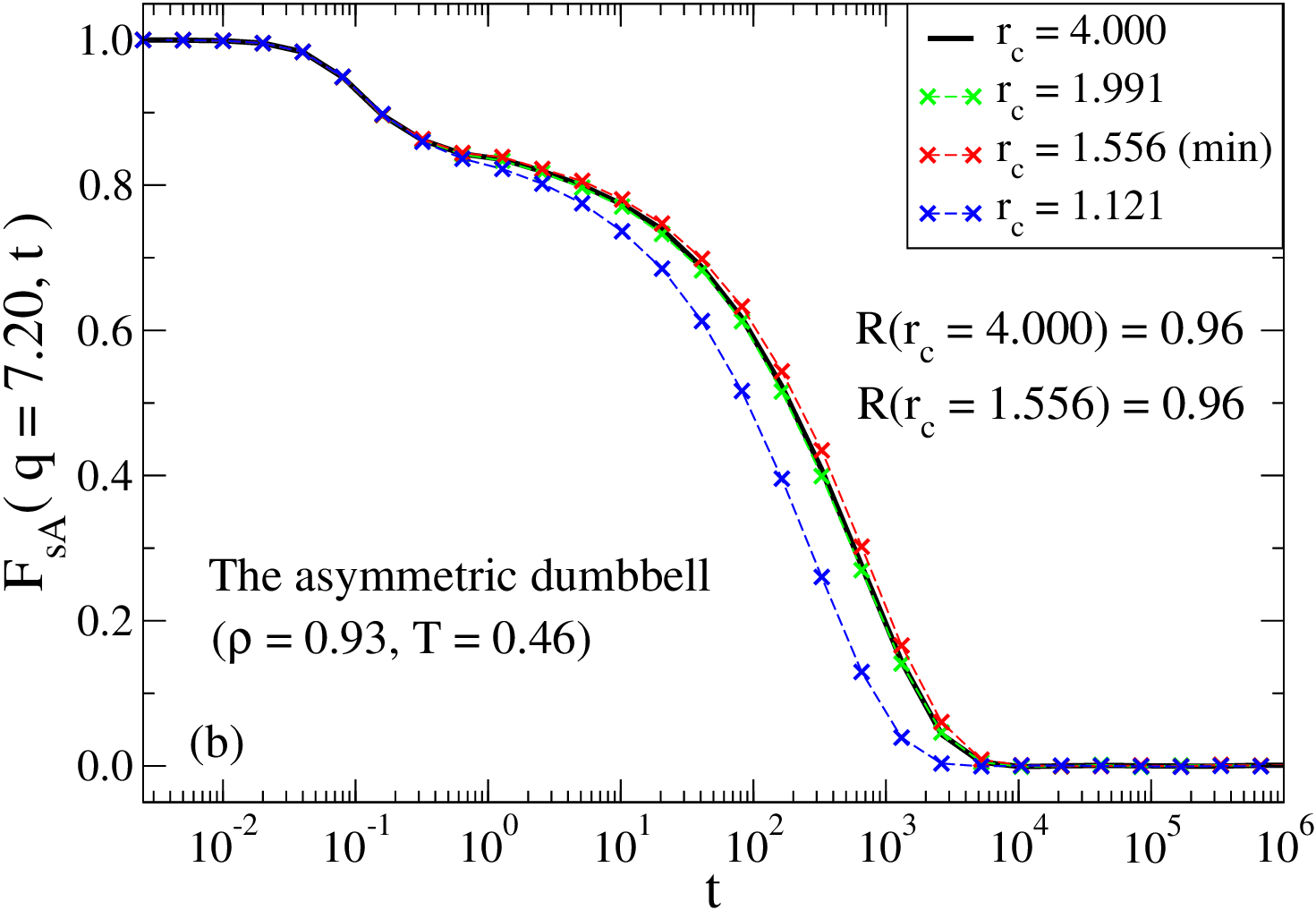}
  \caption{The effect on structure and dynamics of varying the cutoff for the asymmetric dumbbell model.
    The red and black curves give, respectively, results for an FCS cutoff and a large reference cutoff. Note that in this case the FCS cutoff is defined by using the {\it second} minimum (the first minimum is not the absolute minimum).
    (a) RDF at $\rho=0.93$ and $T = 0.46$ ($R=0.96$). The inset quantifies the deviation in RDF from the reference RDF (black curve) as  a function of the cutoff. 
(b) A-particle {ISF} at the same state point.}
  \label{Asym}
\end{figure}

Figures \ref{OTP}(a) and (b) show the LJ particle RDF and ISF of the OTP model. Both quantities are well approximated using an FCS cutoff, although slight deviations are noted for the ISF (red curve, see Appendix B for considerations concerning this). The OTP model is a border-line strongly correlating liquid ($R=0.91$).
\begin{figure}[H]
  \centering
  \includegraphics[width=70mm]{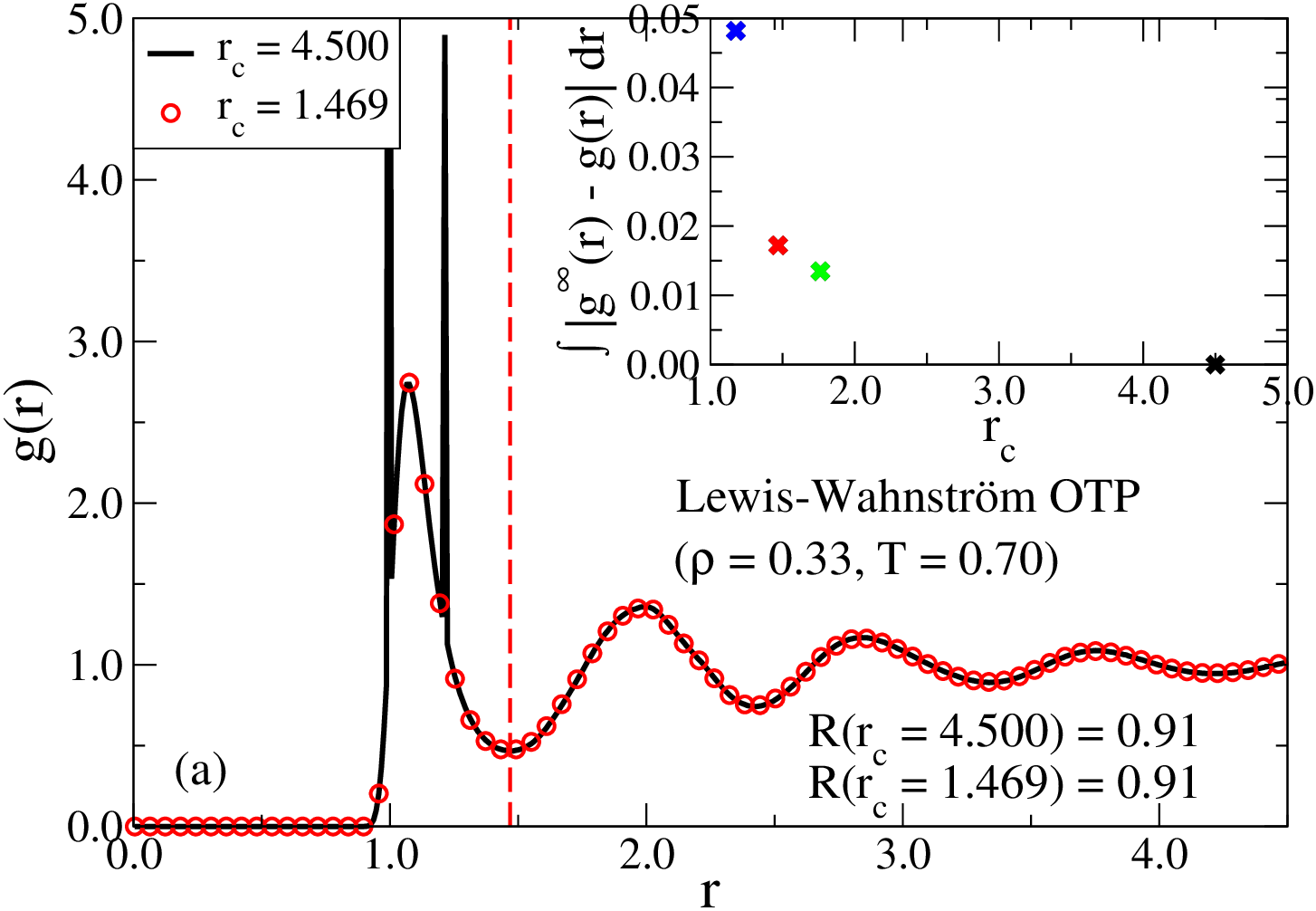}
  \vspace{-20pt}
\end{figure}
Figures \ref{Asym}(a) and (b) show corresponding figures for the large (A) particle of the asymmetric dumbbell model at a viscous state point. The use of an FCS cutoff gives accurate results for both structure and dynamics. The FCS cutoff was placed at the {\it second} minimum of the AA RDF, because the {AA RDF} has here a lower value than at the first minimum. If the cutoff is placed at the first minimum, clear deviations are noted (data not shown).

\begin{figure}[H]
  \centering
  \includegraphics[width=70mm]{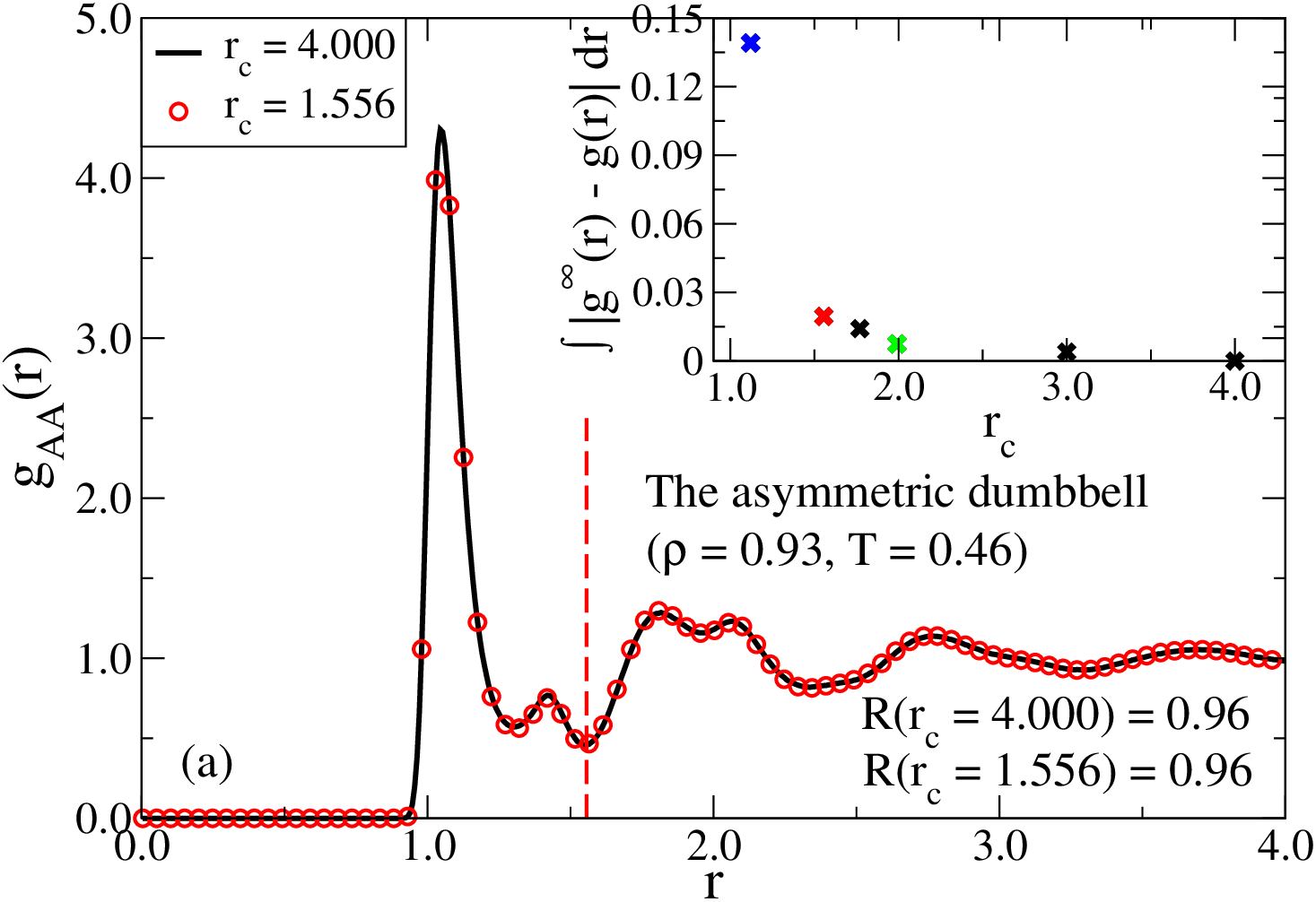}
  \vspace{-20pt}
\end{figure}

\subsection{Rigid {SPC/E} water}

\begin{figure}[H]
  \centering
  \includegraphics[width=70mm]{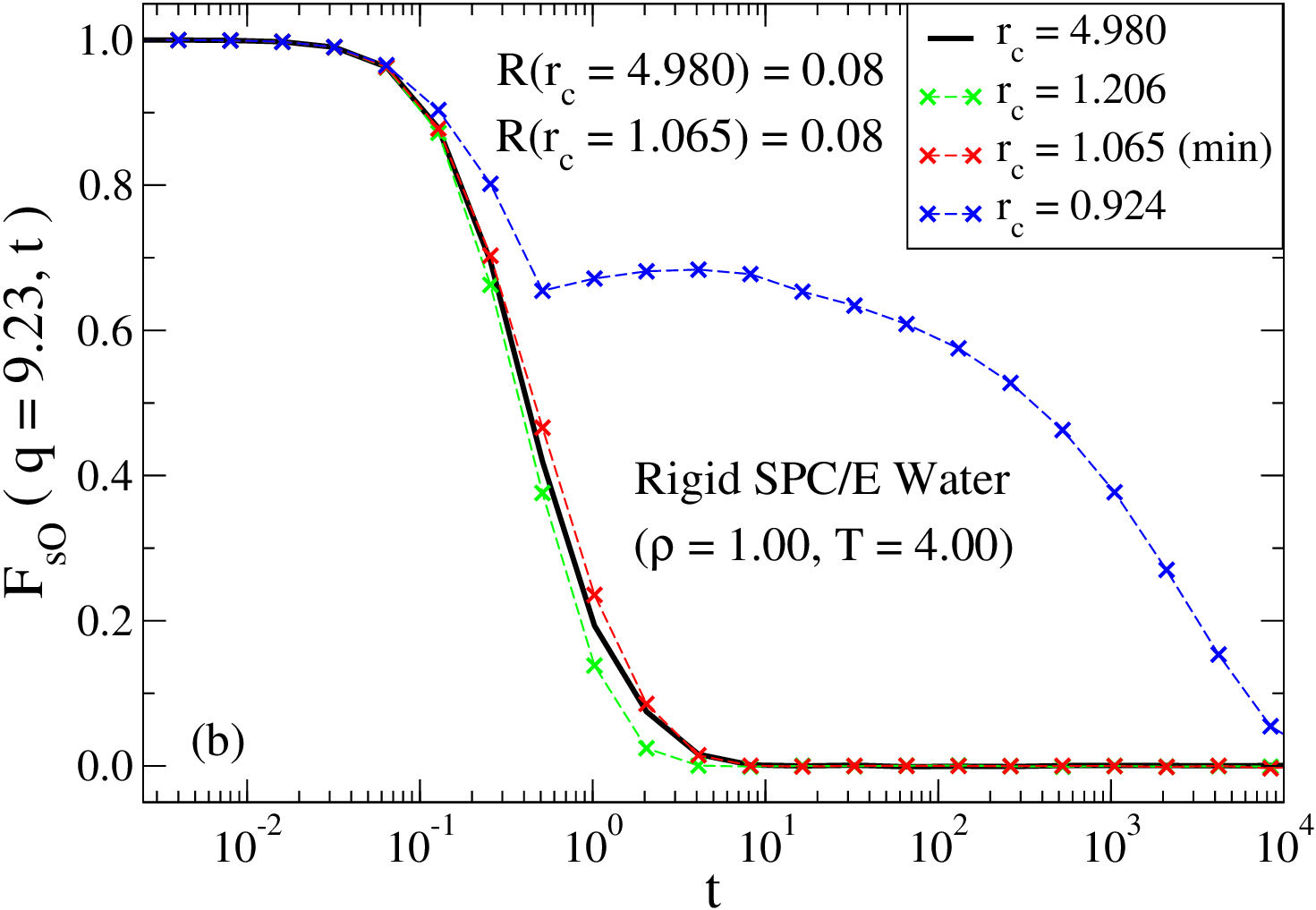}
  \caption{The effect on structure and dynamics of varying the cutoff for the rigid {SPC/E} water model \cite{spce}.
    The red and black curves give, respectively, results for an FCS and a large reference cutoff.
    (a) Oxygen-oxygen RDF at $\rho=1.00$ and $T = 4.00$ ($R=0.08$). The inset quantifies the deviation in RDF from the reference RDF as a function of the cutoff. 
(b) Oxygen {ISF} at the same state point.}
  \label{water}
\end{figure}

We consider finally the rigid {SPC/E} water model \cite{spce} (Fig. \ref{water}). This model is not strongly correlating at ambient conditions, a fact that reflects water's well-known density maximum \cite{paperII}. The structure of the {SPC/E} water model is not well represented using an FCS cutoff. The FCS cutoff dynamics on the other hand shows only slight deviations from that of the reference curve (black).

\begin{figure}[H]
  \centering
  \includegraphics[width=70mm]{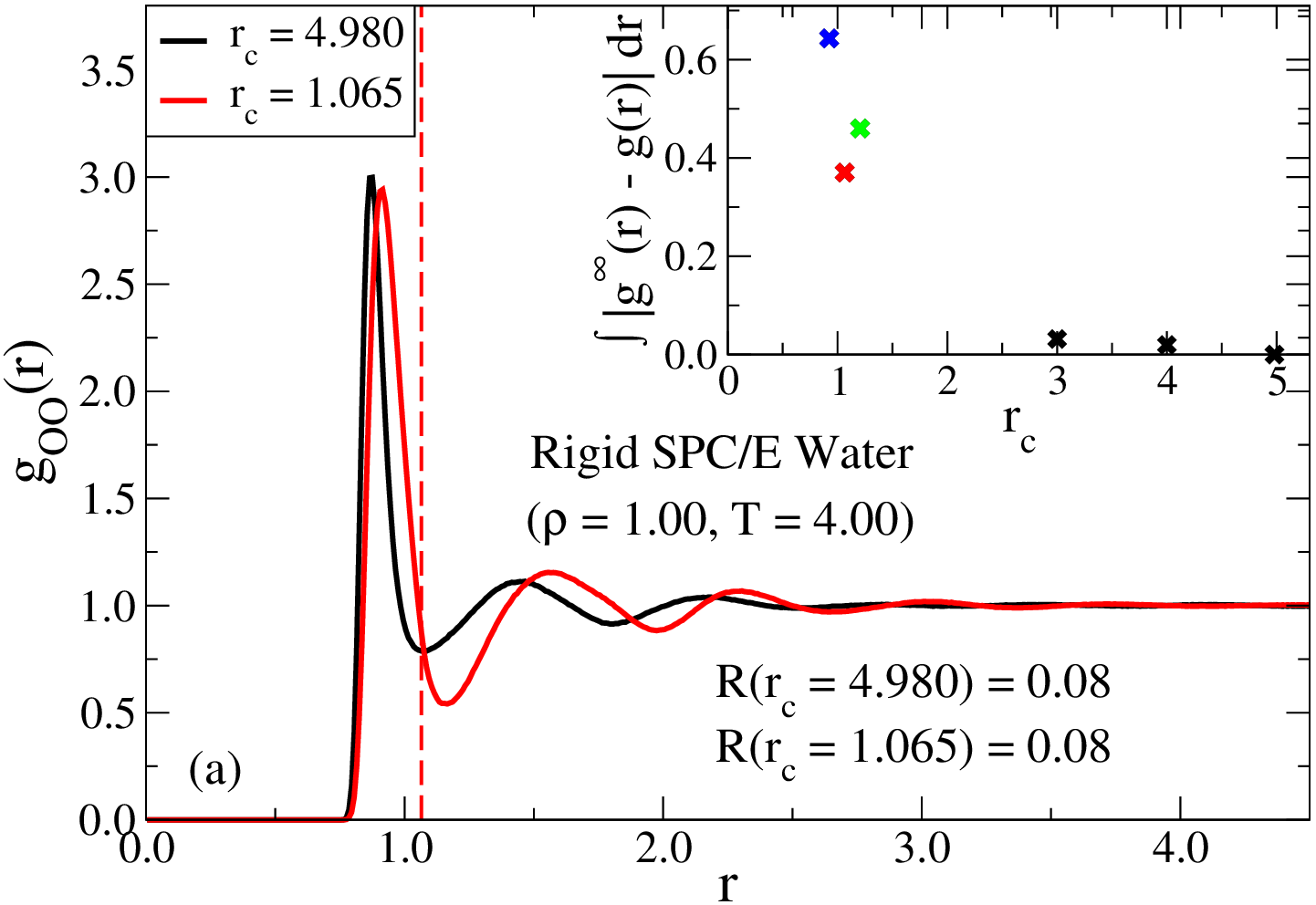}
  \vspace{-20pt}
\end{figure}

\section{Summarizing the simulation results}\label{sum}

The previous section showed that structure and dynamics are well approximated in simulations using an FCS cutoff for the following atomic and molecular systems:

\begin{itemize}
\item Inverse power-law (IPL) systems ($n=18, 6, 4$),
\item single-component Lennard-Jones liquid at density $\rho=0.85$,
\item generalized Kob-Andersen binary Lennard-Jones mixtures,
\item Wahnstr{\"o}m binary Lennard-Jones mixture,
\item single-component Buckingham liquid,
\item Lewis-Wahnstr{\"o}m OTP model,
\item asymmetric dumbbell model.
\end{itemize}
These systems are all strongly correlating \cite{paperI,paperII,paperIII,paperIV,paperV}. Thus for strongly correlating liquids it is enough to know the intermolecular interactions within the FCS in order to accurately simulate structure and dynamics. 

The simulations showed further that for all of the following atomic and molecular systems structure and/or dynamics are not properly reproduced when an FCS cutoff is used:
\begin{itemize}
\item Dzugutov (DZ) liquid,
\item Lennard-Jones Gaussian (LJG) liquid,
\item Gaussian core model (GCM),
\item Hansen-McDonald molten salt model,
\item rigid {SPC/E} water model.
\end{itemize}
None of these liquids are strongly correlating. For all these systems larger cutoffs are needed in order to faithfully reproduce the system's physics. 

In conclusion, a shifted-forces FCS cutoff leads to accurate results {\it if and only if} the liquid is strongly correlating at the state point in question. We know of no exceptions to this rule. This suggests that  strongly correlating liquids are {\it characterized} by the property that intermolecular interactions beyond the FCS can be ignored.

\section{The real essence of liquid simplicity}\label{SCL_sec}

As mentioned in the introduction a definition of simple liquids is most useful if it identifies their {\it real essence} \cite{locke}, the underlying fundamental characteristic from which these liquids' simple features, their {\it nominal essences}, follow. We suggest below that the class of simple liquids is to be identified with the class of strongly correlating liquids (Sec. \ref{sldef}). This is motivated by first summarizing the many simple properties of strongly correlating liquids (Sec. \ref{prop}), then showing that this class of liquids can be characterized from three different perspectives: mathematically, physically, and chemically (Sec. \ref{eqv}). This gives three very different but equivalent  characterizations, indicating that the class of strongly correlating liquids is fundamental and further motivating the suggestion that the real essence of liquid simplicity is the existence of strong correlations of virial/potential-energy equilibrium $NVT$ fluctuations. By connecting to the chemists' concept of non-associated liquids we then discuss which real-world liquids are simple (Sec. \ref{simple}), discuss briefly liquids near interfaces (Sec. \ref{interface}), and give examples of complex liquid properties (Sec. \ref{complex}). Finally, Sec. \ref{perturbation} points out that our results call into question traditional perturbation theory, which is based on assuming quite different roles of the attractive and the repulsive forces.

\subsection{Strongly correlating liquids and their simple properties}\label{prop}

Most properties of strongly correlating liquids follow from the existence of  ``isomorphs'' in their phase diagram (see below). Some of their properties were identified before isomorphs were defined in 2009 \cite{paperIV}, however, for instance that
\begin{itemize}
\item all eight fundamental thermoviscoelastic response functions are given in terms of just one, i.e., the dynamic Prigogine-Defay ratio is close to unity \cite{baileyOrderParameters},
\item aging may be described by adding merely one extra parameter \cite{paperII,paperIII},
\item power-law density scaling \cite{rol05} is obeyed to a good approximation, i.e., for varying density and temperature the relaxation time is a function of $\rho^\gamma/T$ \cite{sch09}.
\end{itemize}

An isomorph is an equivalence class of state points in a system's phase diagram. Two state points ($\rho_{1}$, $T_{1}$) and ($\rho_{2}$, $T_{2}$) are defined to be isomorphic if the following holds \cite{paperIV}: Whenever one microconfiguration of state point ($1$) and one of state point ($2$) have the same reduced coordinates (i.e., $\rho_{1}^{1/3} \textbf{r}^{(1)}_{i} = \rho_{2}^{1/3} \textbf{r}^{(2)}_{i}$ for all particles $i$), these two microconfigurations have proportional configurational Boltzmann factors,

\begin{equation} \label{defiso}
  e^{-U(\textbf{r}_ {1}^{(1)}, ..., \textbf{r}_ {N}^{(1)})/k_{B}T_{1}} = C_{12}e^{-U(\textbf{r}_{1}^{(2)}, ...,  \textbf{r}_ {N}^{(2)})/k_{B}T_{2}}.
\end{equation}
For most systems the isomorph concept is approximate just as $WU$ correlations are rarely perfect. Thus we do not require Eq. (\ref{defiso}) to be rigorously obeyed for all microconfigurations, but only to a good approximation and only for all {\it physically relevant} microconfigurations. By this is meant microconfigurations that are not {\it a priori} unimportant for the physics. Being an equivalence class, an isomorph defines a continuous curve of state points in the liquid's phase diagram. Only liquids for which the potential energy is an Euler homogeneous function, e.g., systems with IPL pair potentials, have exact isomorphs. An IPL fluid with $v(r)\propto r^{-n}$ has exact isomorphs characterized by $\rho^{\gamma} /T = {\rm Const.}$ where $\gamma = n/3$. 

Appendix A of Ref. \onlinecite{paperIV} showed that a liquid is strongly correlating if and only if it has isomorphs to a good approximation. This was confirmed in Refs. \onlinecite{paperIV} and \onlinecite{paperV}, which showed that Lennard-Jones type atomic liquids have good isomorphs.  Reference \onlinecite{ing11} showed that the strongly correlating Lewis-Wahnstr{\"o}m OTP and asymmetric dumbbell molecular models also have good isomorphs.

Equation (\ref{defiso}) has many consequences. These were derived and discussed in detail in the original isomorph paper from 2009 (Ref. \onlinecite{paperIV}), to which the reader is referred. Basically, structure and dynamics at two isomorphic state points are identical in reduced units. Quantities that are invariant along an isomorph include (but are not limited to):

\begin{enumerate}
\item The excess entropy, i.e., the entropy in excess of the ideal gas entropy at the same density and temperature -- this is the configurational contribution to the entropy (a term that is negative because a liquid is always more ordered than an ideal gas at same density and temperature).
\item All $N$-body entropy terms. Recall that the excess entropy can be expanded in a series of two-body, three-body, etc, terms. Each term is invariant along an isomorph \cite{paperIV}.
\item The isochoric heat capacity. 
\item The structure in reduced units (defined by $\tilde{\textbf{r}}_{i} \equiv \rho^{1/3}\textbf{r}_{i}$ for all particles $i$). Not only the radial distribution function, but all higher-order distribution functions in reduced units are isomorph invariant.
\item The Newtonian {\it NVE} and Nos$\acute{e}$-Hoover {\it NVT} equations of motion in reduced units; likewise Brownian dynamics.
\item All autocorrelation functions in reduced units.
\item All average relaxation times in reduced units.
\item Reduced transport coefficients like the diffusion coefficient, viscosity, etc.
\end{enumerate}
Isomorphs have the further interesting property that there is no relaxation for an instantaneous change of temperature and density when jumping from an equilibrated state point to a different state point isomorphic with the initial state. The absence of relaxation derives from the fact that the Boltzmann probabilities of scaled microconfigurations are identical. Such ``isomorph jumps'' work quite well for the KABLJ liquid \cite{paperIV}, for the asymmetric dumbbell, and for the Lewis-Wahnstr{\"o}m OTP molecular models \cite{ing11}. Moreover, the effective temperature of a glass prepared by a temperature-density jump from an equilibrium state of a strongly correlating liquid depends only on the final density \cite{effectivetemperature}; this provides yet another example of a simple feature of these liquids.

Some further predictions for the class of strongly correlating liquids deriving from the existence of isomorphs are:

\begin{itemize}
\item The solid-liquid coexistence curve is an isomorph \cite{paperIV,paperV}. This implies invariance along the coexistence curve of the reduced structure factor, the reduced viscosity, the reduced diffusion constant, etc, as well as pressure invariance of the melting entropy, and a reduced-unit Lindemann melting criterion \cite{paperIV}.
\item Collapse of the two-order-parameter maps of Debenedetti {\it et al.} \cite{order1,order2,order3,order4,order5} to one-dimensional curves \cite{paperIV}. 
\item Isochronal superposition \cite{nga05}, i.e., the fact that when pressure and temperature are varied, the average relaxation time determines the entire relaxation spectrum \cite{paperIV}. 
\end{itemize}

The above listed properties of strongly correlating liquids all reflect {\it simple} features of strongly correlating liquids. A final, recently established simple property is a thermodynamic separation identity: for all strongly correlating liquids, if $s$ is the excess entropy per particle, the temperature as a function of $s$ and density $\rho$ factorizes as follows \cite{ing11a}

\begin{equation}\label{eos}
T\,=\,f(s)h(\rho)\,.
\end{equation}
Equation (\ref{eos}) has a number of consequences \cite{ing11a}, including the Gruneisen equation of state and that the isomorphs of LJ liquids -- in particular, the LJ solid-liquid coexistence curve -- are given by $(A\rho^4-B\rho^2)/T={\rm Const.}$ \cite{khrapak,beyond}.

\subsection{Mathematical, physical, and chemical characterizations of strongly correlating liquids}\label{eqv}

At a given state point, if the average potential energy is denoted by $\langle U\rangle$, the constant-potential-energy hypersurface is defined by $\Omega=\{(\br_1,...,\br_N)\in R^{3N}|U(\br_1,...,\br_N)=\langle U\rangle\}$. This is a compact, Riemannian $(3N-1)$-dimensional differentiable manifold. Each state point has its own such hypersurface. In this way, a family of manifolds is defined throughout the phase diagram. In Appendix A of Ref. \onlinecite{paperIV} it was shown that the reduced-unit constant-potential-energy manifold is invariant along a strongly correlating liquid's isomorphs, and that invariance curves exist for these manifolds only for strongly correlating liquids. Thus, for such liquids these manifolds constitute a one-parameter family of manifolds, not two-parameter families as expected from the fact that the phase diagram is two-dimensional. 

The above provides a {\it mathematical} characterization of the class of strongly correlating liquids. The {\it physical} characterization was discussed already: the existence of isomorphs in the phase diagram. A liquid is strongly correlating if and only if it has isomorphs to a good approximation (Appendix A of Ref. \onlinecite{paperIV}).  

The {\it chemical} characterization of strongly correlating liquids is the property documented in the present paper: A liquid is strongly correlating at a given state point if and only if the liquid's structure and dynamics are accurately calculated by shifted-forces cutoff simulations that ignore interactions beyond the first coordination shell. This is an empirical finding for which we have at present no compelling arguments, but can at present merely point out two things. First, the property of insignificance of interactions beyond the FCS is an isomorph invariant: {\it If} a liquid has good isomorphs and {\it if} an FCS cutoff works well at one state point, FCS cutoffs must work well for all its isomorphic state points. Thus the chemical characterization of strongly correlating liquids is consistent with the fact that these liquids have isomorphs. Secondly, almost all of the fluctuations in virial and potential energy of the LJ liquid come from interparticle separations within the FCS \cite{paperII}. 

The new ``FCS characterization'' of strongly correlating liquids shows that these liquids are characterized by having a well-defined FCS. Most likely it is the existence of a well-defined FCS which implies the almost cancellation of the linear term of the shifted-force potential. The fact that interactions beyond the FCS may be ignored shows that interactions are effectively short ranged, which means that the structure is dominated by what may be termed packing effects.

\subsection{Defining the class of ``simple liquids''}\label{sldef}

Section \ref{prop} listed strongly correlating liquids' many simple properties. Section \ref{eqv} showed that this liquid class may be characterized from three quite different points of view. Clearly the class of strongly correlating liquids is fundamental. Since the properties of strongly correlating liquids are generally simpler than those of liquids in general, we now propose the definition

\begin{itemize}
\item {\bf Simple liquids = Strongly correlating liquids}
\end{itemize}
This is the basic message of the present paper, which implies a quantification of the degree of simplicity via the number  $R$ of Eq. (\ref{Rdef}), the $NVT$ ensemble equilibrium virial/potential-energy correlation coefficient.

Compared to the standard definition of simple liquids (as those with radially symmetric pair interactions) there are some notable differences: 

\begin{enumerate}
\item Simplicity is quantified by a continuously variable, it is not an on/off property.
\item The degree of simplicity generally varies throughout the phase diagram. Most strongly correlating liquids loose this property as the critical point is approached; on the other hand many or most non-strongly correlating liquids are expected to become strongly correlating at very high pressures \cite{highpressurescl}.
\item Not all ``atomic'' liquids (i.e., with radially symmetric pair interactions) have simple regions in the low-pressure part of the phase diagram (compare the Dzugutov, Lennard-Jones Gaussian, Gaussian core, and molten salt models).
\item Not all simple liquids are atomic (compare the Wahnstr{\"o}m OTP and the asymmetric dumbbell models). 
\end{enumerate}

According to the new definition of liquid simplicity the case where the potential energy is an Euler homogeneous function of the particle positions ($R=1$) sets the gold standard for simplicity. This is consistent with the many simple properties of IPL liquids. Due to the absence of attractions, IPL fluids have no liquid-gas phase transition. In this sense it may seem strange to claim that IPL fluids are the simplest liquids. However, more realistic strongly correlating liquids like the LJ liquid cease to be so when the liquid-vapor coexistence line is approached near the critical point, showing that this phase transition cannot be understood in the framework of simple liquids. This contrasts with the liquid-solid phase transition, where for instance the fact that the coexistence line for simple liquids is an isomorph -- confirmed by simulations of the LJ liquid \cite{paperV} -- explains several previously noted regularities \cite{paperIV,khrapak}.

Is the hard-sphere fluid simple? One may define a configurational virial function for this system, but it is not obvious how to define a potential energy function that is different from zero. Thus there is no meaningful correlation coefficient $R$ for hard-sphere fluids. On the other hand, the hard-sphere liquid may be regarded as the $n\rightarrow\infty$ limit of an IPL liquid, and it is well known that for instance the hard-sphere radial distribution function is close to that of, e.g., an $r^{-20}$ IPL liquid at a suitably chosen temperature. This would indicate that hard-sphere liquids are simple, consistent with the prevailing point of view. Another interesting case is that of the Weeks-Chandler-Andersen (WCA) version of the LJ liquid, which cuts off all attractions by putting the force equal to zero beyond the potential energy minimum. This liquid is strongly correlating \cite{cos09}, but in simulations that the WCALJ liquid has somewhat poorer isomorphs than the LJ liquid. 

It is possible that the hard-sphere liquid and the WCALJ liquid should be both excluded from the class of simple liquids on the grounds that their potentials are not analytic. For systems interacting with pair potentials, for instance, it could make good sense to add the extra requirement that the pair potential is an analytical function of the inverse pair distance, i.e., that an expansion exists of the form $v(r)=\sum_n v_n r^{-n}$. Such an extra analyticity requirement would not exclude any strongly correlating liquids occurring in nature where all potentials are expected to be analytic.

\subsection{Which liquids in the real world are simple?}\label{simple}

Real-world liquids may be classified according to the nature of the chemical bonds between the molecules. There are five types of bonds \cite{pauling}, listed below with a few typical examples (polymeric systems may be added as a separate class): 

\begin{itemize}
\item Van der Waals bonds (argon, toluene, butane, ...); 
\item Metallic bonds (gold, aluminum, alloys, ...); 
\item Hydrogen bonds (water, glycerol, ethanol, ...); 
\item Ionic bonds (molten sodium chloride and potassium nitrate, room-temperature ionic liquids, ...); 
\item Covalent bonds (silica and borate melts, ...).
\end{itemize}
Most liquids involve elements of more than one type of chemical bonds. For instance, van der Waals forces are present in all liquids; the first class consists merely of those liquids that {\rm only} have van der Waals forces. Another borderline example is a dipolar organic liquid like di-butyl-phthalate, where van der Waals as well as Coulomb forces are present; a liquid like glycerol has also strong dipolar interactions, i.e., an element of the ionic bonds that defines class 4), etc.

Based on computer simulations and known properties of liquids we believe that most or all van der Waals and metallic liquids are strongly correlating \cite{ped08,baileyOrderParameters,paperII}, i.e., {\it simple}. Liquids that are not simple are the hydrogen-, ionically, and covalently bonding liquids. In these cases the virial/potential-energy correlations are weakened by one or the other form of competing interactions.

Metals play a special role as simple liquids, because their interatomic forces derive from collective interactions between ion cores and free electrons \cite{fis64}. The resulting interaction is a non-directional interaction between symmetric ion cores, i.e., these systems are ``simple'' in the traditional sense. Preliminary computer simulations of ours show that metals are strongly correlating \cite{paperI}, so metals are simple also in the sense of this paper, as well. However, not all isomorph invariants apply for metals. For instance, the electron gas can influence the collective dynamics without any visible structural and relaxational counterpart \cite{bov01,pet11}, so isomorph invariance most likely breaks down for these (fast) collective degrees of freedom.

It should be emphasized that the above considerations refer to ambient or moderate pressure conditions. It was recently suggested that if crystallization is avoided, all liquids become strongly correlating at sufficiently high pressure \cite{highpressurescl}.  Thus, e.g., the molten silicates of the earth's upper mantle are predicted to be simpler than molten silicates at ambient pressure.

\subsection{Liquids near interfaces}\label{interface}

Since the property of being simple cannot be read of from knowledge of the potential alone, it is of interest to consider liquids under more general circumstances, for instance under confinement or generally near interfaces. Liquids near interfaces show rich and complicated behavior. For instance, a liquid confined to the nanoscale may change its dynamic properties several orders of magnitude compared to the bulk system. Predicting these changes is an important challenge relevant for biological systems, engineered devices, etc. Recently, it was shown that some liquids retain bulk liquid behavior in confinement 
\cite{mit06,LJconfinedsmoothwalls,connectiondynamicsprofile,dumbbellconfinedroughwalls}. More specifically, it was shown that Rosenfeld's excess entropy scaling in the bulk persists in confinement and is to a good approximation independent of the wall-fluid interaction strength and the degree of confinement. This was shown for LJ and hard-sphere liquids, suggesting the possibility of extending the concept of a simple liquid to apply beyond bulk systems. Given that interactions in strongly correlating liquids are limited in range by the radius of the FCS, one may speculate that these liquids are simple also by having the property that an external field at one point affects the liquid over shorter distances than for a general liquid. Clearly, more work is needed to clarify the relevance and consequences of the present definition of liquid simplicity near interfaces and in external fields \cite{wee97,wee02}.

\subsection{A note on complex liquid behavior}\label{complex}

Liquids that are not simple in the above defined sense of the term often have complex properties \cite{order3,SCS1,SCS2,SCS3,tetraAnomalies}. Water with its correlation coefficient close to zero at ambient conditions is a prime example of a complex liquid. It is well known for water that a certain region of state points in the density/temperature phase diagram exhibits anomalous thermodynamic behavior in the sense that isobaric heating implies densification. Numerical evidence indicates that these state points lie within a larger region with diffusion anomaly, i.e., an increased diffusivity upon isothermal compression \cite{order3}, a region that in turn lies within a larger region of structural anomaly characterized by decreasing order upon isothermal compression \cite{order3}. 

Different order parameters exist for characterizing the structural order of liquids, some of which relate purely to an integral over the RDF \cite{SCS1,SCS2,SCS3}. In this way it is possible to calculate the contribution to structural anomalies from the different coordination shells \cite{SCS1,SCS2,SCS3}. It has been shown \cite{SCS2,SCS3} that the structural anomaly of water and waterlike liquids is not a: ''{... first-shell effect. Rather, they reflect how structuring in second and more distant coordination shells responds to changes in thermodynamic or system parameters.}'' The full anomalous behavior of water derives from interactions beyond the FCS \cite{SCS1,SCS3}. This is consistent with the results presented in this paper, since the structure and dynamics of strongly correlating liquids are given exclusively by the interactions within the FCS.

\subsection{To which extent do the assumptions of standard pertubation theory hold?}\label{perturbation}

The finding that the FCS plays a crucial role for a large class of systems may be taken as a modern demonstration of the classic van der Waals picture of liquids in the sense that such liquids can be understood in terms of packing effects \cite{row82}. On the other hand, our results call into question the basis of traditional perturbation theory, which is also usually traced back to van der Waals \cite{wid67}. Perturbation theory is based on the assumption of entirely different roles being played by the repulsive and the attractive forces \cite{wid67,wca,bar76,han05,row82,bar76,wee97,zho09}: The repulsive forces define the structure and reduce the entropy compared to that of an ideal gas at same density and temperature, the attractive forces reduce the pressure and energy compared to that of an ideal gas. From the findings of the present and a previous paper \cite{FCS2} it is clear, however, that this picture applies only when the FCS coincides roughly with the pair potential minimum, i.e., at fairly low pressure. At very high pressures the whole FCS is within the range of the repulsive forces, here the attractive forces play little role for simple liquids. In general, what is important for a strongly correlating liquid is to take into account properly all forces from particles within the FCS -- and only these. Thus the well-known WCA reference system, which ignores the attractions, is a good reference only at such high pressures that all forces from particles within the FCS are repulsive \cite{wcabertier1,wcabertier2,FCS2}. 

The dominance of the FCS for simple liquids reflects the fundamental physics that the characteristic length defining the potential minimum (e.g., $\sigma$ of the LJ potential) is much less important than generally believed: $\sigma$ determines the density of the low-pressure condensed phase, but that is all. The only physically important length for simple liquids is that given by the macroscopic density itself: $\rho^{-1/3}$. At low pressures this length is roughly that of the potential energy minimum, explaining why the latter has been generally assumed to be important.

The above considerations apply only for simple liquids; in general both lengths play important roles for the physics. The irrelevance of any length defined by the microscopic potential emphasizes once again that the class of strongly correlating liquids is at the one end of the ``complexity scale'' where, at the other end, one finds systems like macromolecules, electrolytes, interfaces, micelles, or enzymes, for which multiple length and time scales are important \cite{bag10}.

\section{Concluding remarks}\label{disc}

If you ask a chemist what is a simple liquid, he or she may answer that nonassociated liquids are simple, whereas associated liquids are much more complex. These two concepts are defined as follows in Chandler's textbook \cite{chandler}. The intermolecular structure of a {\it nonassociated} liquid ``can be understood in terms of packing. There are no highly specific interactions in these systems.'' In contrast, water is an example of an {\it associated} liquid, and its ``linear hydrogen bonding tends to produce a local tetrahedral ordering that is distinct from what would be predicted by only considering the size and shape of the molecule'' \cite{chandler}. 

Packing usually refers to purely entropic, hard-sphere like behavior. Given that no realistic potentials are infinitely repulsive it makes good sense to interpret packing more generally as all short-ranged effects of the intermolecular interactions. If one accepts this more general interpretation of packing, the crucial role of the FCS for strongly correlating liquids is consistent with the understanding since long that the properties of nonassociated liquids can be interpreted in terms of packing:
\begin{itemize}
\item Once the forces from particles within the FCS are known, basically everything is known. 
\end{itemize}
In other words, for a simple liquid there are no important long-range interactions, and ``considering the size and shape of the molecule'' is enough to account for the liquid's physical properties. This applies even for the $r^{-4}$ IPL fluid, which has a fairly long-ranged interaction.

The present definition of the class of simple liquids is thus consistent with the chemists' general picture of simple liquids. The new definition goes further, however, by quantifying simplicity via the virial/potential-energy correlation coefficient $R$ of Eq. (\ref{Rdef}). In particular, the degree of simplicity is not an on/off property of the potential, but varies continuously with state point. Thus even a complex liquid like water is expected to approach simple behavior under sufficiently high pressure \cite{highpressurescl} and, conversely, the prototype strongly correlating LJ liquid becomes gradually more complex as density is lowered and the critical region and/or the gas phase is approached. Is this a problem, given that everyone agrees that the gas phase is simple? We do not think so. In fact, the gas phase is simple for a different reason, namely that molecules move freely most of the time, only interrupted by occasional fast and violent collisions with other molecules. It would be strange if a system exhibiting one form of simplicity could be transformed continuously, while maintaining its simplicity, into a system of an entirely different form of simplicity; one would expect the intermediate phase to be complicated.

Liquid simplicity is characterized by the correlation coefficient $R$ of Eq. (\ref{Rdef}) being close to unity, i.e., that $1-R$ is a small number. This situation is typical in physics, where simplifying features always appear when some  dimensionless number is small. The obvious question arises whether a perturbation theory may be constructed around simple liquids, embracing the more complex ones. Only future can tell whether this is possible, but it does present a challenge because the properties of IPL fluids ($R=1$) cannot be worked out analytically.

A potentially annoying feature about defining liquid simplicity from the existence of strong correlations of the virial/potential-energy fluctuations is that one cannot read off directly from potential and state point whether or not a given liquid is simple. The same holds for the existence of isomorphs or the property that an FCS cutoff reproduces the correct physics. We believe one should accept this as an acceptable cost for precisely delimiting the simple liquids from others. With the power of today's computers this is much less of a problem than previously. For most systems a brief simulation will determine whether or not the liquid is strongly correlating at the state point in question. 

Except for the IPL fluids, no system is simple in the entire fluid phase. This paper focused on the condensed liquid phase, not too far from the solid-liquid coexistence line, but far from the critical point and the gas phase -- it is here that some liquids are simple. The focus on liquids is not meant to imply a limitation to the liquid phase, however. In fact, simulations show that when a strongly correlating liquid crystallizes, the crystal is at least as strongly correlating \cite{paperII}. A  theory has been developed for (classical) strongly correlating crystals, showing that the property of strong virial/potential-energy equilibrium fluctuations in the $NVT$ ensemble is an anharmonic effect that survives as $T\rightarrow 0$ \cite{paperII}. Of course, low-temperature crystals are not classical systems, and both for liquids and crystals an interesting topic for future works is the implication of the proposed simplicity definition for the quantum description of condensed matter.

Section \ref{prop} summarized the several nominal essences of simple liquids. What is the {\it real essence} of liquid simplicity? Given that three fundamental characterizations of strongly correlating liquids are equivalent -- the mathematical, the physical, and the new chemical (FCS) characterization -- this question cannot be answered unequivocally. At the end of the day it is a matter of taste whether one defines liquid simplicity from the existence of strong virial/potential-energy correlations, from the existence of isomorphs, from the existence of invariance curves for constant-potential-energy hypersurfaces, or from the property that interactions beyond the FCS play little role. All four properties are equivalent.

\acknowledgments

The centre for viscous liquid dynamics ``Glass and Time'' is sponsored by the Danish National Research Foundation (DNRF). We gratefully acknowledge useful inputs from Livia Bove, Jesper Schmidt Hansen, and S{\o}ren Toxv{\ae}rd.

\appendix

\section{Model details}\label{pairP}

The model systems investigated are listed below. Quantities are given in rationalized units defined by setting $\epsilon=\sigma=1$. Masses that are not specified are unity.

\noindent{\it Single-component inverse power-law (IPL) fluids}: $N=1024$ particles interacting via 
$v(r) = \epsilon ( \sigma/r)^{n}$. Three different fluids were studied ($n=18,6,4$). 

\noindent{\it Single-component Lennard-Jones liquid}: $N= 1024$ particles interacting via Eq. (\ref{LJ}).

\noindent{\it Generalized Kob-Andersen binary mixture} \cite{ka1,ka2}: A binary mixture of 820 $A$ particles and 204 $B$ particles interacting via 

  $v(r) = \epsilon_{\alpha \beta} /(12-n) \left[ n( \sigma_{\alpha \beta} /r)^{12} - 12 ( \sigma_{\alpha \beta} /r)^{n}\right]$.
Binary mixtures with $n=4, 10$ were studied. The parameters used are: $\epsilon_{AA} = 1.00$, $\epsilon_{AB}=1.50$, $\epsilon_{BB}=0.50$,  $\sigma_{AA}=2^{1/6}$, $\sigma_{AB}=0.8\cdot2^{1/6}$, $\sigma_{BB}=0.88\cdot2^{1/6}$.

\noindent{\it Buckingham liquid}: $N=1000$ particles interacting via 
$v(r) = \epsilon \left[ 6 / (\alpha - 6) \exp[\alpha(1- r/r_{m})] - \alpha /(\alpha -6)(r_{m}/r)^{6}\right]$.
The parameters used are $\epsilon=1$, $\alpha=14.5$, $r_{m}=2^{1/6}$.

\noindent{\it Dzugutov liquid} \cite{dzugutov}: $N=1024$ particles interacting via 
$v(r) = v_1 + v_2$ where $v_1 = \left( A ( r^{-n} - B )\exp(c/(r - a)) \right)$ and $v_2= B \exp( d/(r - b))$  and $r \geq a \Rightarrow v_{1} = 0$,  $ r \geq b \Rightarrow v_{2} = 0$ ($a<b$).
The parameters used are $a=1.87$, $b=1.94$, $c=1.1$, $d=0.27$, $A=5.82$, $B=1.28$, $n=16$.

\noindent{\it Lennard-Jones Gaussian liquid} \cite{LJG}: $N=1024$ particles interacting via 
$v(r) = \epsilon \left( (\sigma/r)^{12} -  2(\sigma/r)^{6} + \epsilon_{0} \exp[- (r - r_0)^2/2 \sigma_{0}^{2}] \right)$ 
The parameters used are $\sigma_{0}^2=0.02$, $\epsilon_{0}=1.50$, $r_{0}=1.47$. 
  
\noindent{\it Gaussian core model} \cite{LJG}: $N=1024$ particles interacting via $v(r) = \epsilon \exp\left[-(r/\sigma)^2\right]$.

\noindent{\it The Hansen-McDonald molten salt model} \cite{moltensalt}: $N=2744$ particles forming an equimolar binary mixture of singly charged cations and anions. The  potential between two particles of charge $q_\alpha$ and $q_\beta$ is given by  
$v(r) = (1/9)r^{-9} + q_{\alpha}q_{\beta}/r$,  where $q_+=1$, $q_{-}=-1$. 

\noindent{\it Lewis-Wahnstr{\"o}m OTP} \cite{otp1,otp2}: This model consists of three identical LJ particles rigidly bonded in an isosceles triangle with unity sides and top-angle of $75^\circ$ (number of molecules studied: $N=320$). 

\noindent{\it The asymmetric dumbbell model} \cite{sch09}: This model consists of a large (A) and a small (B) LJ particle, rigidly bonded with bond distance $r_{AB} = 0.29/0.4963$ (number of molecules studied: $500$). The asymmetric dumbbell model has $\sigma_{BB}=0.3910/0.4963$, $\epsilon_{BB}=0.66944/5.726$, and $m_{B}=15.035/77.106$. The $A-B$ interaction between different molecules is determined by the Lorentz-Berthelot mixing rule.

\noindent{\it Rigid SPC/E water} \cite{spce}: The water model is an isosceles triangle with sides $r_{OH} = 1/3.166$ and base line $2r_{OH}*\sin(109.47/2)$ (number of molecules studied: $1000$). The oxygen-oxygen intermolecular interactions are given by the LJ pair potential ($\epsilon_{OO}=1$, $\sigma_{OO}=1$, and $m_{O}=15.9994/1.00794$). There are no intermolecular LJ interactions for $H$-$H$ or $H$-$O$. The three particles are charged with $q_{O}=-0.8476e/(4\pi\varepsilon_{0}3.166${\AA}$0.650kJ/mol)^{1/2}$ and $q_{H}=|q_{O}|/2$ ensuring charge neutrality.

\section{How to delimit the first coordination shell?}\label{FCS_sec}

\begin{figure}[H]
  \centering
  \includegraphics[width=70mm]{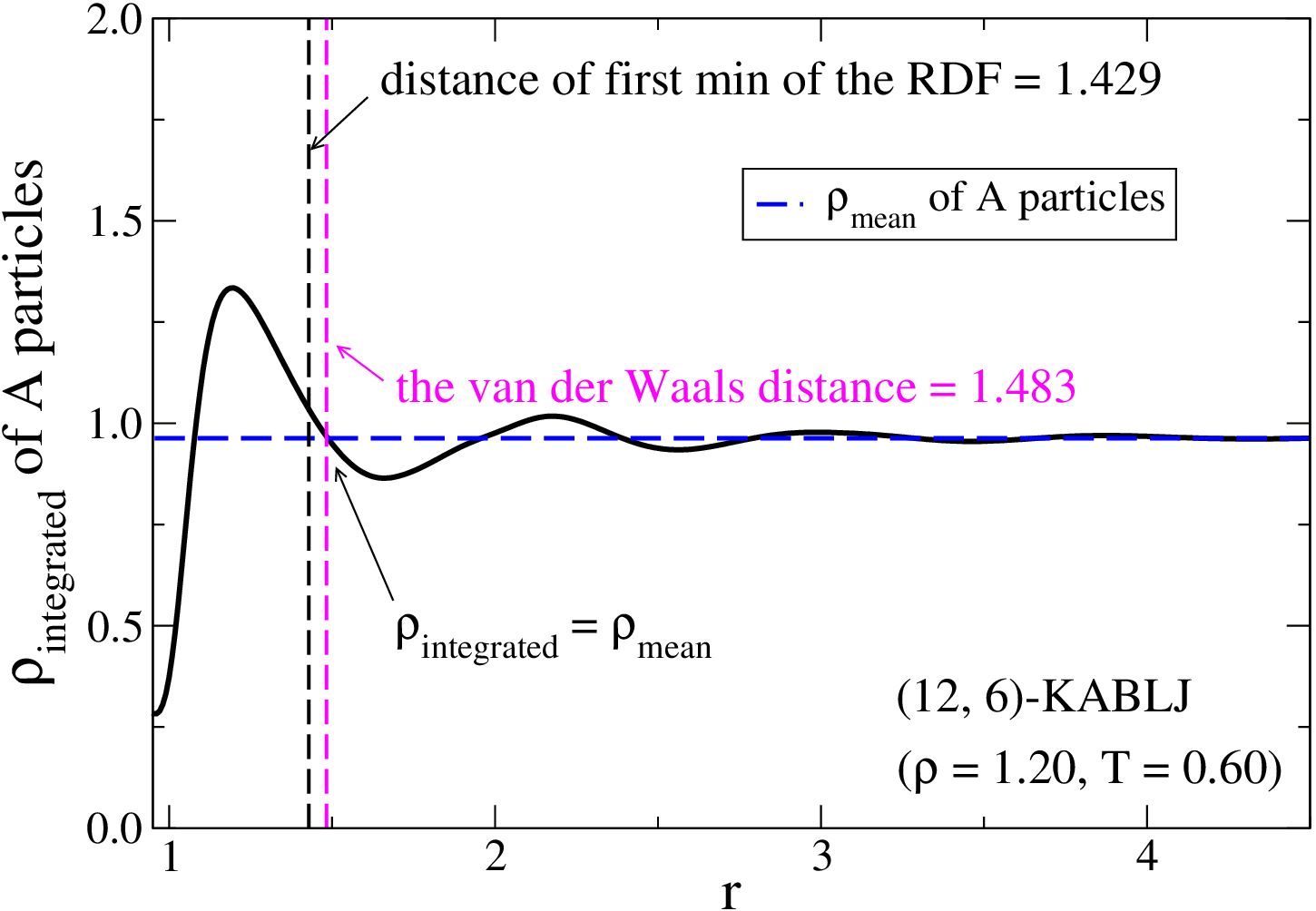}
  \caption{Integrated local density $\rho_{integrated}$ of the A particles of the standard KABLJ mixture plotted as a function of distance to the origin (calculated from the A-particle RDF). The dashed black line gives the distance of the minimum of the RDF, the dashed magenta line gives the van der Waals distance, i.e., the distance at which the integrated local density equals the overall average density of the system. The horizontal dashed blue line marks the mean density $\rho_{\rm mean}$ of $A$ particles in the simulated system ($\rho_{mean} = 0.8 * 1.2)$.}
  \label{meanDensity}
\end{figure}

In all simulations the FCS cutoff was defined by placing the cutoff at the first minimum of the RDF, which is the standard definition of the FCS for liquids \cite{chandler}. An alternative definition goes back to van der Waals \cite{row82}. The FCS is here identified with a sphere of radius determined by requiring that the average density within the FCS $\rho_{integrated}$ equals the overall average density $\rho_{\rm mean}$. Figure \ref{meanDensity} shows for the KABLJ system the integrated local density of $A$ particles calculated from the RDF (including the particle at the center) as a function of the distance to the origin. The ``van der Waals distance'' is slightly larger than the first minimum of the RDF.    
 
\begin{figure}[H]
  \centering
  \includegraphics[width=70mm]{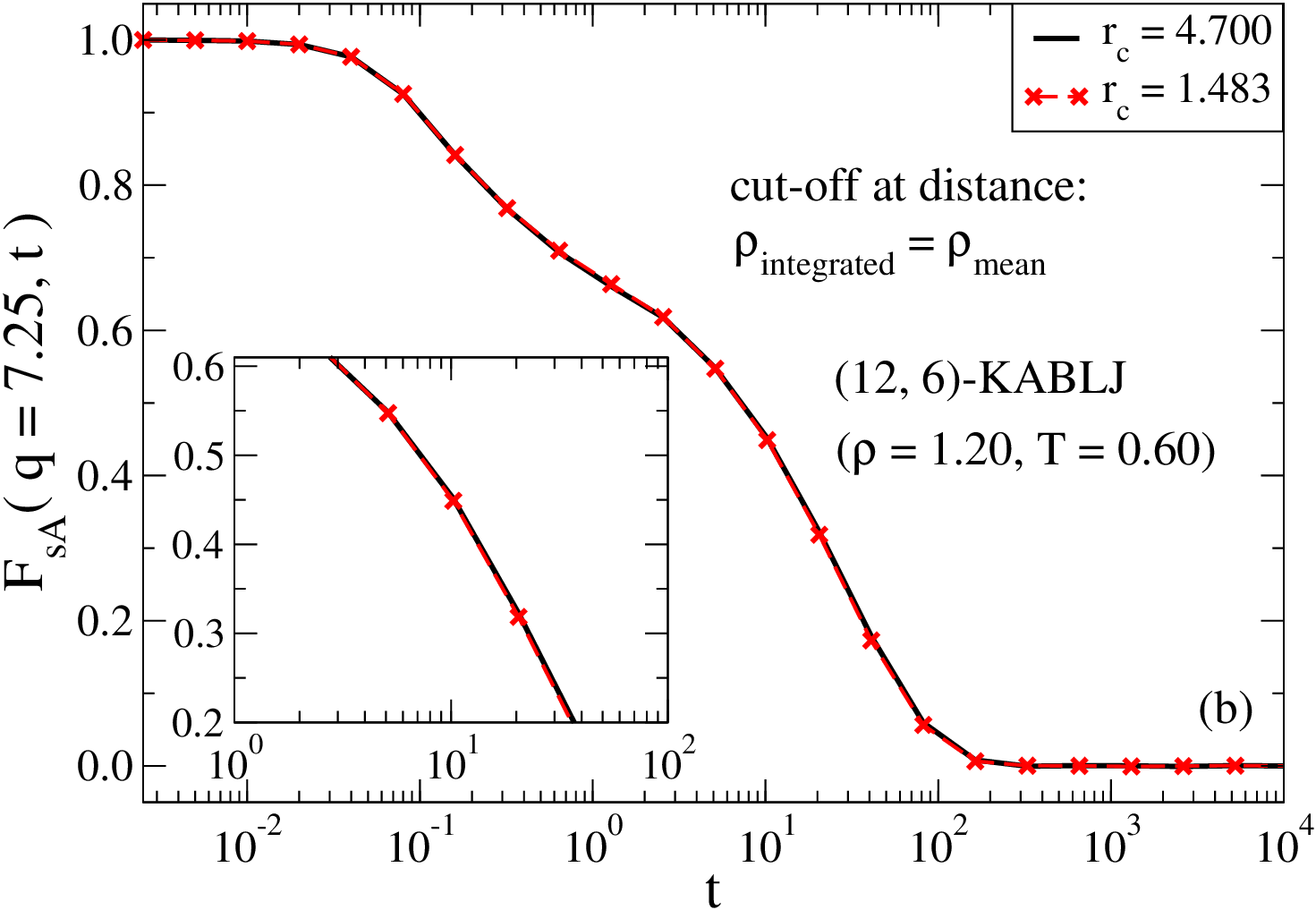}
  \caption{Effect on the $A$-particle ISF for the standard KABLJ mixture of different of ways of delimiting the FCS. 
(a)  FCS is identified from the minimum beyond the first peak of the $AA$-particle RDF.
(b)  FCS is identified by the van der Waals distance, i.e., the distance at which the integrated local density equals the mean density of the system. The van der Waals distance is slightly larger than the RDF minimum. Using the latter as defining the FCS cutoff radius gives a better representation of the dynamics, as seen from the inset.}
  \label{sizeFCS}
\end{figure}

We applied this alternative definition of the FCS cutoff in Fig. \ref{sizeFCS}, which shows the $A$-particle {ISF}  for the ($12$, $6$)-KABLJ mixture of Fig. \ref{meanDensity} simulated with, respectively, a cutoff at the first minimum of the RDF (Fig. \ref{sizeFCS}(a)) and the van der Waals cutoff (Fig. \ref{sizeFCS}(b)).

\begin{figure}[H]
  \centering
  \includegraphics[width=70mm]{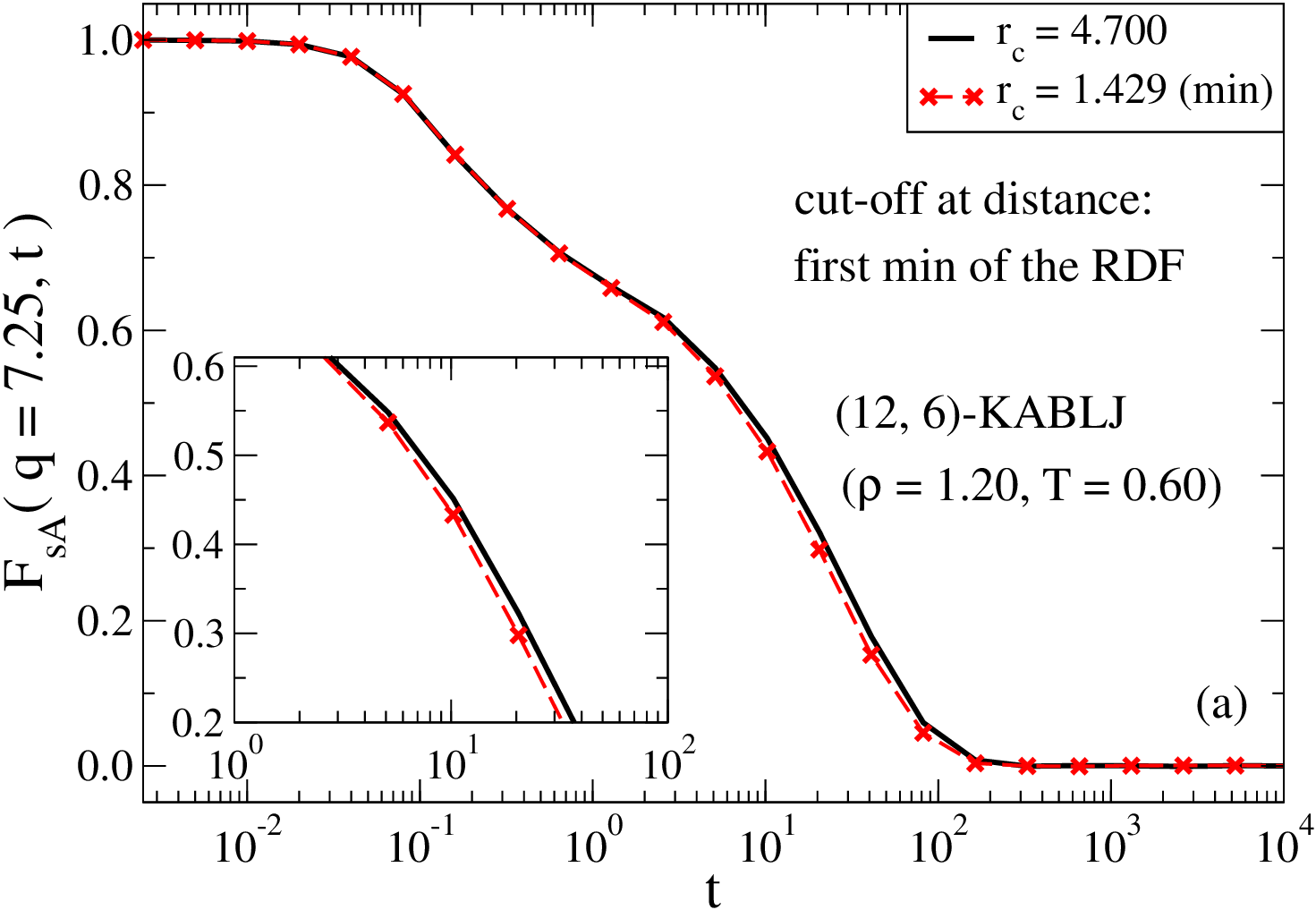}
  \vspace{-20pt}
\end{figure}    
Although the cutoff difference is merely 0.05, the van der Waals cutoff approximates better the reference ISF than does the RDF minimum cutoff. Thus it is possible that the van der Waals distance may serve as a better definition of the size of the FCS than the standard FCS definition.

\begin{figure}[H]
  \centering
  \includegraphics[width=70mm]{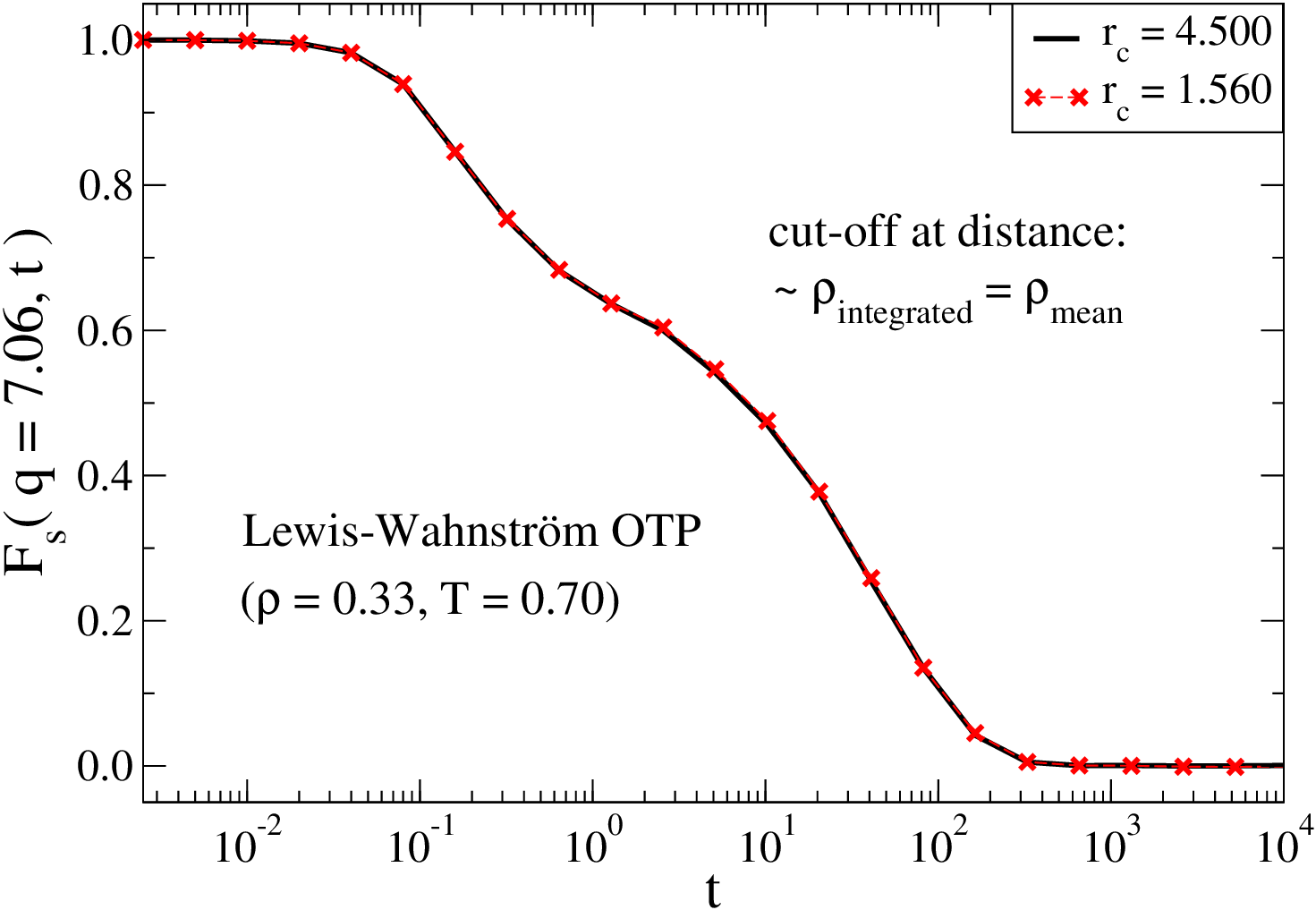}
  \caption{Results for the ISF of the Lewis-Wahnstr{\"o}m OTP model with a cutoff at $r_{c}=1.56$ (red), and a large reference cutoff (black). The deviations of Fig. \ref{OTP}(b), in which $r_c=1.47$, disappear by choosing this slightly larger cutoff, not far from the van der Waals cutoff ($r_c=1.53$).}
  \label{sizeOTP}
\end{figure}

Identifying the size of the FCS for molecular systems is less straightforward, especially when different intermolecular interactions are involved. It is noteworthy how well the simple cutoff scheme in Fig. \ref{Asym} represents the dynamics of the asymmetric dumbbell model. The slight deviations observed for the OTP model (Fig. \ref{OTP}(b)), disappear when the cutoff is increased from $r_c=1.47$ to $r_{c} = 1.56$ (Fig. \ref{sizeOTP}). This distance is close, but not identical, to the van der Waals distance calculated from the particle RDF ($\approx 1.53$). More work is needed to clarify the best way to delimit the FCS and define the FCS cutoff.

\bibliography{mybib}

\end{document}